\documentclass[aps,prc,preprint,amsmath,amssymb,showpacs,preprintnumbers,superscriptaddress]{revtex4-1}
\usepackage{CJK}
\usepackage{graphicx}
\usepackage{dcolumn}
\usepackage{bm}
\usepackage{color}
\usepackage{hyperref}

\allowdisplaybreaks[4]

\begin{document}

\title{Time-dependent covariant density functional theory in 3D lattice space: benchmark calculation for ${}^{16}{\rm O}+{}^{16}{\rm O}$ reaction}
\author{Z. X. Ren}
\affiliation{State Key Laboratory of Nuclear Physics and Technology, School of Physics, Peking University, Beijing 100871, China}

\author{P. W. Zhao}
\email{pwzhao@pku.edu.cn}
\affiliation{State Key Laboratory of Nuclear Physics and Technology, School of Physics, Peking University, Beijing 100871, China}

\author{J. Meng}
\email{mengj@pku.edu.cn}
\affiliation{State Key Laboratory of Nuclear Physics and Technology, School of Physics, Peking University, Beijing 100871, China}

\begin{abstract}
  Time-dependent covariant density functional theory with the successful density functional PC-PK1 is developed in a three-dimensional coordinate space without any symmetry restrictions, and benchmark calculations for the ${}^{16}{\rm O}+{}^{16}{\rm O}$ reaction are performed systematically.
  The relativistic kinematics, the conservation laws of the momentum, total energy, and particle number, as well as the time-reversal invariance are examined and confirmed to be satisfied numerically.
  Two primary applications including the dissipation dynamics and above-barrier fusion cross sections are illustrated.
  The obtained results are in good agreement with the ones given by the nonrelativistic time-dependent density functional theory and the data available.
  This demonstrates that the newly developed time-dependent covariant density functional theory could serve as an effective approach for the future studies of nuclear  dynamical processes.
\end{abstract}


\maketitle

\section{Introduction}
During the past decades, new experimental facilities with radioactive beams have extended our knowledge of nuclear chart to the very limits of nuclear binding, in particular to the unstable neutron-rich nuclei.
Many novel and striking features have been found in the structure of neutron-rich nuclei, such as the halo phenomenon, and the disappearance of traditional magic numbers and occurrence of new ones~\cite{Tanihata2013PPNP}.
The new observations do not only provide us new insights for nuclear systems, but also challenge the established nuclear theory.

Enormous efforts have been made to understand the physics of nuclear many-body systems based on microscopic approaches.
The nuclear density functional theory (DFT) is one of the most popular approaches in this context~\cite{Bender2003Self, meng2016relativistic}.
Starting from a universal energy density functional, the complicated nuclear many-body problem can be simplified as a one-body problem~\cite{Kohn1965DFT}.
In this way, the DFT can provide a global description for almost all nuclei in the nuclear chart including very neutron-rich nuclei, and a fairly good accuracy has been achieved with only a few parameters in the energy density functional.

By taking into account the Lorentz symmetry, the covariant density functional theory (CDFT) has attracted a lot of attention in nuclear physics~\cite{RING1996PPNP,Vretenar2005PhysicsReport,meng2006PPNP,NIKSIC2011PPNP,meng2016relativistic}.
In this framework, the nucleons are treated as Dirac particles moving in large scalar and vector fields with the order of a few hundred MeV~\cite{Volum16}.
This brings many advantages to describe the nuclear systems with the CDFT, such as the new saturation mechanism of nuclear matter~\cite{walecka1974theory}, the natural inclusion of spin-orbit interactions~\cite{Sharma1995Pb_shift} and, thus, the relativistic spin and pseudospin symmetries~\cite{liang2015hidden}.
Another important advantage of the CDFT is the self-consistent treatment of the time-odd fields, which share the same coupling constants as the time-even ones thanks to the Lorentz invariance~\cite{Vretenar2005PhysicsReport, Meng2013FT_TAC}.
With these advantages, CDFT has been successfully used to investigate the ground-state properties of many exotic nuclei~\cite{meng1996relativistic, meng1998giant, zhou2010neutron, xia2018ADNDT} and also various nuclear excitation phenomena including rotations~\cite{Peng2008maganetic_roration, Zhao2011PRL_AMR, Zhao2015Rod-shaped, Zhao2017ChiralRotation} and vibrations~\cite{Niksic2002DDME1_QRPA, Paar2007RPP, Paar2009QRPA, Niu2009FTQRPA}.

The time-dependent DFT (TDDFT) is a dynamical extension of DFT~\cite{Rung1984TDDFT} for describing dynamical processes of many-body systems.
In nuclear physics, the development of TDDFT can be traced back to the mid 1970s~\cite{ENGEL1975215NPA, Bonche19761DTDHF, KOONIN1976TDHF_O16,Cusson1976TDHFO16, Koonin1977TDHFO16, Flocard1978TDHFO16, Bonche1978TDHFO16, Davies1978TDHFO16}, which are known under the notation of the time-dependent Hartree-Fock method~\cite{dirac1930TDHF}.
However, the early applications of the nuclear TDDFT were suffered from the simplified effective interactions and/or restricted geometric symmetries~\cite{Negele1982TDDFT}.
With the ever-improving computational capabilities, the TDDFT experienced a revival during the last twenty years, and the unrestricted three-dimensional (3D) calculations with modern nuclear density functionals become available~\cite{Simene2012PEPJA, NakatsukasaRMP2016, SIMENEL2018TDHF_PPNP, STEVENSON2019PPNP}.
Up to now, the TDDFT in 3D lattice space has been widely applied to many nuclear dynamical processes, such as the multinucleon transfer process~\cite{Simenel2010MNT, Sekizawa2013TDHFMNT, Sekizawa2016TDHF_Ni_U, Wu2019MNT},
fission~\cite{Goddard2015TDHFfission, Bulgac2016Pu240_fission, Tanimura2017fission, scamps2018impact},
fusion~\cite{Guo2012fusion,Umar2015SHE_TDHF, Yu2017TDHF3D, Guo2018TDHF_fusion, Guo2018tensor_fusion},
collective vibration~\cite{maruhn2005TDHF_GDR, Reinhard2007TDHF_GR, Schuetrumpf2016TABC},
cluster scattering~\cite{Umar2010TDHF_C12}, etc.

The dynamical extension of the CDFT, i.e., the time-dependent CDFT (TDCDFT), can be traced back to the early 1980s,
where the time-dependent versions of the Walecka model were adopted to describe the dynamics of colliding nuclear slabs~\cite{MULLER1981TDRMF} and relativistic heavy ion collisions~\cite{Cusson1985TDCDFT, Bai1987TDCDFT}.
Later on, the time-dependent relativistic mean-field theory is used to describe the dynamics of Coulomb excitations of nuclei by assuming axial symmetry~\cite{Vretenar1993TDRMF, VRETENAR1995TDRMF}.
In the present work, TDCDFT with the successful density functional PC-PK1 is developed in a three-dimensional coordinate space without any symmetry restrictions.
This would be helpful to clarify the ambiguity of the spin-orbit fields and time-odd fields in the nonrelativistic TDDFTs and, thus, provide a new framework to investigate the dynamical processes of nuclei.
However, such a development is not simple at all because of the longstanding difficulties in solving the CDFT in a 3D lattice~\cite{zhang2009first, ZhangIJMPE2010}.
Recently, the CDFT has been solved in a 3D lattice space with the inverse Hamiltonian~\cite{hagino2010iterative, tanimura20153d} and Fourier spectral methods~\cite{REN2017Dirac3D}, and its successful applications includes the studies of nuclear linear-chain ~\cite{Ren2019C12LCS} and toroidal structures~\cite{REN2020Toroidal}.
This paves the way to develop the corresponding time-dependent approaches in a full 3D lattice space without assuming any symmetries.

In our very recent work~\cite{Ren2020HeBeTDCDFT}, the TDCDFT was developed in a 3D lattice space with relativistic density functionals and applied to investigate the microscopic dynamics of the linear-chain cluster states.
Following the previous work, a systematic investigation of the ${}^{16}{\rm O}+{}^{16}{\rm O}$ reaction will be reported in this work with the detailed formalism of the TDCDFT in 3D lattice space.
In Sec.~\ref{sec_theo}, the theoretical framework is introduced.
The numerical details are given in Sec.~\ref{sec_nume}.
Section~\ref{sec_numericaltest} is devoted to the numerical tests.
Two primary applications, including the dissipation dynamics and above-barrier fusion cross sections, are presented in Secs.~\ref{sec_DisDy} and \ref{sec_fusion}, respectively.
Finally, a summary is given in Sec.~\ref{sec_summ}.

\section{Theoretical framework}\label{sec_theo}
\subsection{Covariant density functional theory}
The starting point of the CDFT is a standard Lagrangian density which, in the point-coupling form, can be written as~\cite{ZhaoPC-PK1},
\begin{equation}
  \begin{split}
    \mathcal{L}=\,&\mathcal{L}^{\rm free}+\mathcal{L}^{\rm 4f}+\mathcal{L}^{\rm hot}+\mathcal{L}^{\rm der}+\mathcal{L}^{\rm em}\\
    =\,&\bar{\psi}(i\gamma^\mu\partial_\mu-m_N)\psi-\frac{1}{2}\alpha_S(\bar{\psi}\psi)(\bar{\psi}\psi)-\frac{1}{2}\alpha_V(\bar{\psi}\gamma^\mu\psi)(\bar{\psi}\gamma_\mu\psi)-\frac{1}{2}\alpha_{TV}(\bar{\psi}\vec{\tau}\gamma^\mu\psi)\cdot(\bar{\psi}\vec{\tau}\gamma_\mu\psi)\\
    &-\frac{1}{3}\beta_S(\bar{\psi}\psi)^3-\frac{1}{4}\gamma_S(\bar{\psi}\psi)^4-\frac{1}{4}\gamma_V[(\bar{\psi}\gamma^\mu\psi)(\bar{\psi}\gamma_\mu\psi)]^2-\frac{1}{2}\delta_S\partial^\nu(\bar{\psi}\psi)\partial_\nu(\bar{\psi}\psi)\\
    &-\frac{1}{2}\delta_V\partial^\nu(\bar{\psi}\gamma^\mu\psi)\partial_\nu(\bar{\psi}\gamma_\mu\psi)-\frac{1}{2}\delta_{TV}\partial^\nu(\bar{\psi}\vec{\tau}\gamma^\mu\psi)\cdot\partial_\nu(\bar{\psi}\vec{\tau}\gamma_\mu\psi)\\
    &-\frac{1}{4}F^{\mu\nu}F_{\mu\nu}-e\frac{1-\tau_3}{2}(\bar{\psi}\gamma^\mu\psi)A_\mu.
  \end{split}
\end{equation}
It includes the Lagrangian density for free nucleons $\mathcal{L}^{\rm free}$,
the four-fermion point-coupling terms $\mathcal{L}^{\rm 4f}$, the higher-order terms $\mathcal{L}^{\rm hot}$ accounting for the medium effects,
the derivative terms $\mathcal{L}^{\rm der}$ to simulate the finite-range effects that are crucial for a quantitative description of nuclear density distributions,
and the electromagnetic interaction terms $\mathcal{L}^{\rm em}$.
Thus, one can build the energy density functional for a nuclear system,
\begin{equation}\label{Eq_energy_functional}
  \begin{split}
    E_{\rm tot}=\,&E_{\rm kin}+E_{\rm int}+E_{\rm em}\\
    =\,&\int d^3r~\left\{\sum_{k=1}^A\psi_k^\dag(\bm{\alpha}\cdot\hat{\bm{p}}+\beta m_N)\psi_k+\frac{1}{2}\alpha_S\rho_S^2+\frac{1}{3}\beta_S\rho_S^3+\frac{1}{4}\gamma_S\rho_S^4+\frac{1}{2}\delta_S\rho_S\Delta\rho_S\right.\\
    &+\frac{1}{2}\alpha_Vj^\mu j_\mu+\frac{1}{4}\gamma_V(j^\mu j_\mu)^2+\frac{1}{2}\delta_V j^\mu\Delta j_\mu+\frac{1}{2}\alpha_{TV}j^\mu_{TV}(j_{TV})_\mu+\frac{1}{2}\delta_{TV}j_{TV}^\mu\Delta(j_{TV})_\mu\\
    &+\left.ej_c^\mu A_\mu+\frac{1}{2}A_\mu\Delta A^\mu\right\},
  \end{split}
\end{equation}
where $E_{\rm kin}$, $E_{\rm int}$, and $E_{\rm em}$ are the kinetic, interaction, and electromagnetic energies, respectively.
The local densities and currents $\rho_S$, $j^\mu$, $j_{TV}^\mu$, and $j_c^\mu$ are given by,
\begin{subequations}\label{Eq_density_current}
  \begin{align}
    &\rho_s=\sum_{k=1}^A\bar{\psi}_k\psi_k,\\
    &j^\mu=\sum_{k=1}^A\bar{\psi}_k\gamma^\mu\psi_k,\\
    &j_{TV}^\mu=\sum_{k=1}^A\bar{\psi}_k\gamma_\mu\tau_3\psi_k,\\
    &j_c^\mu=\sum_{k=1}^A\bar{\psi}_i\gamma^\mu\frac{1-\tau_3}{2}\psi_k,
  \end{align}
\end{subequations}
where $\tau_3$ is the isospin Pauli matrix with the eigenvalues $+1$ for neutrons and $-1$ for protons.
The time component $j^0$ is usually denoted as the vector density $\rho_v$.

In the static case, the densities and currents in Eq.~\eqref{Eq_density_current} are time-independent.
By means of the variation of energy density functional Eq.~\eqref{Eq_energy_functional} with respect to the densities and currents, one obtains the Kohn-Sham equation for nucleons,
\begin{equation}\label{Eq_static_Dirac_eq}
  \hat{h}(\bm{r})\psi_k(\bm{r})=\varepsilon_k\psi_k(\bm{r}),
\end{equation}
where $\varepsilon_k$ is the single-particle energy and $\hat{h}$ is the single-particle Dirac Hamiltonian,
\begin{equation}\label{Eq_dirac_hamiltonian}
  \hat{h}(\bm{r})=\bm{\alpha}\cdot(\hat{\bm{p}}-\bm{V})+V^0+\beta(m_N+S).
\end{equation}
The scalar $S(\bm{r})$ and four-vector $V^\mu(\bm{r})$ potentials read
\begin{subequations}
  \begin{align}
    S(\bm{r})=\,&\alpha_S\rho_S+\beta_S\rho_S^2+\gamma_S\rho_S^3+\delta_S\Delta\rho_S,\\
    V^\mu(\bm{r})=\,&\alpha_Vj^\mu+\gamma_V(j^\mu j_\mu)j^\mu+\delta_V\Delta j^\mu+\tau_3\alpha_{TV}j_{TV}^\mu+\tau_3\delta_{TV}\Delta j_{TV}^\mu+e\frac{1-\tau_3}{2}A^\mu, \label{Eq_vecpot}
  \end{align}
\end{subequations}
where the electromagnetic field $A^\mu$ is determined by Poisson's equation,
\begin{equation}
   -\Delta A^\mu=ej_c^\mu.
\end{equation}
By solving the Dirac equation Eq.~\eqref{Eq_static_Dirac_eq} self-consistently, one can obtain the single-nucleon wavefunctions for a nucleus in its ground state.

\subsection{Time-dependent covariant density functional theory}
In the dynamical case, the evolution of single-nucleon wavefunctions $\psi_k$ should fulfill the time-dependent Kohn-Sham equation~\cite{Rung1984TDDFT,Leeuwen1999TDDFT},
\begin{equation}\label{Eq_td_Dirac_eq}
  i\frac{\partial}{\partial t}\psi_k(\bm{r},t)=\hat{h}(\bm{r},t)\psi_k(\bm{r},t).
\end{equation}
The time-dependent $\hat{h}(\bm{r},t)$ is purely determined by the time-dependent densities and currents~\cite{Rung1984TDDFT}.
With the \textit{adiabatic approximation}~\cite{NakatsukasaRMP2016}, the time-dependent single-particle Hamiltonian $\hat{h}(\bm{r},t)$ in Eq.~\eqref{Eq_td_Dirac_eq} is taken as the Dirac Hamiltonian in Eq.~\eqref{Eq_dirac_hamiltonian}, in which the ground-state densities and currents Eqs.~\eqref{Eq_density_current} are obtained with the wavefunctions $\psi_k(\bm{r},t)$ at the time $t$.
This obviously lacks the memory effect, i.e., $\hat{h}(\bm{r},t)$ does not depend on the history of the system.

The time-dependent Dirac equation \eqref{Eq_td_Dirac_eq} has the formal solution,
\begin{equation}\label{Eq_formal_sol}
  \psi_k(\bm{r},t)=\hat{\mathcal{T}}\exp\left[-i\int_{t_0}^tdt'~\hat{h}(\bm{r},t')\right]\psi_k(\bm{r},t_0),
\end{equation}
where $\hat{\mathcal{T}}$ represents the time-ordering operation and $t_0$ is the initial time.

For nuclear collisions, the initial wavefunctions $\psi_k(\bm{r},t_0)$ are composed of the single-particle wavefunctions of the two nuclei, which are usually in their ground states, and are obtained from two separate static CDFT calculations.
Subsequently, the two nuclei are placed on the mesh of a 3D lattice space with a large enough distance between them, so that the overlap between their wavefunctions is negligible at the initial time.
Moreover, the nuclei are boosted to set them in motion.

As the Dirac equation is Lorentz covariant, the boost of nuclei can be realized by using the inhomogeneous Lorentz transformation~\cite{greiner2013relativistic}.
Starting from the ground-state single-particle wavefunctions $\psi_k^{(\rm g.s.)}(\bm{r})$, the Lorentz boosted ones $\psi_k'(\bm{r})$ with velocity $\bm{v}$ read,
\begin{equation}\label{Eq_Lorentz_transform}
  \psi_k'(\bm{r})=\hat{S}(\bm{v})\psi_k^{\rm (g.s.)}(\bm{r}')e^{i\varepsilon_k\bm{v}\cdot\bm{r}/\sqrt{1-v^2}},
\end{equation}
where $\hat{S}(\bm{v})$ denotes the transformation on the four components of a Dirac spinor,
\begin{equation}\label{Eq_Lorentz_transform_factor}
  \hat{S}(\bm{v})=\sqrt{\frac{1+\sqrt{1-v^2}}{2\sqrt{1-v^2}}}+[\bm{\alpha}\cdot(\bm{v}/v)]\sqrt{\frac{1-\sqrt{1-v^2}}{2\sqrt{1-v^2}}},
\end{equation}
and $\bm{r}'$ represents the transformed coordinate,
\begin{equation}
  \bm{r}'=\bm{r}+\left(\frac{1}{\sqrt{1-v^2}}-1\right)\frac{(\bm{r}\cdot\bm{v})\bm{v}}{v^2}.
\end{equation}
Note that here the single-particle energy $\varepsilon_k$ is not shifted by the nucleon mass $m_N$.

The Lorentz boost in Eq.~\eqref{Eq_Lorentz_transform} can be connected with the Galilean boost used in the nonrelativistic TDDFT by approaching the nonrelativistic limits [$v/c\approx0$ and $(\varepsilon_k-m_N)/m_N\approx0$], under which the Lorentz boosted wavefunctions in Eq.~\eqref{Eq_Lorentz_transform} become
\begin{equation}
  \psi_k'(\bm{r})\approx\psi_k^{\rm(g.s.)}(\bm{r})e^{im_N\bm{v}\cdot\bm{r}}.
\end{equation}
They are just identical with the Galilean boosted wavefunctions~\cite{Maruhn2014CPC}.

Finally, it should mention that the spatial components of the electromagnetic vector potential $\bm{A}(\bm{r})$ are neglected in the calculations, since their contributions are extremely small.
Although the center-of-mass correction energy is usually included a posteriori in the self-consistent static CDFT calculations,
this strategy is disputable in the time-dependent case.
For instance, it involves only the total mass number and does not account for the masses of the fragments.
Therefore, the center-of-mass correction is neglected in the present TDCDFT calculations.

\section{Numerical details}\label{sec_nume}
In the present work, the density functional PC-PK1~\cite{ZhaoPC-PK1} is employed to study the ${}^{16}{\rm O}+{}^{16}{\rm O}$ reaction.
The Dirac spinors of the nucleons and the potentials are represented in 3D lattice space without any symmetry restriction.
The mesh sizes along the $x$, $y$, and $z$ axes are identical and chosen as $d=0.8$ fm.
The ground state of ${}^{16}{\rm O}$ is calculated in a box with $24\times24\times24$ grid points, while for the time-dependent calculations, a larger box with $30\times30\times50$ grid points is used.
For the initial states of the time-dependent calculations, the centers of the two ${}^{16}{\rm O}$ nuclei are placed in the $z$ axis with a separation distance $16$ fm.
The Poisson equation for the Coulomb potential is solved by the Hockney's method with the isolated boundary condition~\cite{eastwood1979remarks}.

For the numerical implementation of the formal solution \eqref{Eq_formal_sol}, the predictor-corrector strategy~\cite{Maruhn2014CPC} is adopted, in which the evolution time is cut into a series of small time steps $\Delta t$.
Over each time interval $[t,t+\Delta t]$, the single-particle Hamiltonian in Eq.~\eqref{Eq_formal_sol} is approximated as the one at the mid-time $\hat{h}(t+\Delta t/2)$.
Thus, the evolution of the single-particle wavefunction from $t$ to $t+\Delta t$ is obtained as,
\begin{equation}\label{Eq_pc_solution}
  \psi_k(\bm{r},t+\Delta t)\approx\exp\left[-i\hat{h}(\bm{r},t+\Delta t/2)\Delta t\right]\psi_k(\bm{r},t),
\end{equation}
which also provides the initial condition for the evolution over $[t+\Delta t,t+2\Delta t]$.

In this work, the single-particle Hamiltonian $\hat{h}(t+\Delta t/2)$ is determined with a two-step recipe, i.e., first roughly constructed and then corrected to be a better one.
In the first step, the densities and currents at time $t+\Delta t$, denoted generally as $\tilde{\rho}^{(1)}(t+\Delta t)$, are estimated from $\tilde{\psi}_k^{(1)}(\bm{r},t+\Delta t)$,
\begin{equation}
  \tilde{\psi}_k^{(1)}(\bm{r},t+\Delta t) = \exp\left[-i\hat{h}(\bm{r},t)\Delta t\right]\psi_k(\bm{r},t).
\end{equation}
The Hamiltonian $\hat{h}^{(1)}(\bm{r},t+\Delta t/2)$ is roughly constructed using the average densities and currents $[\rho(\bm{r},t)+\tilde{\rho}^{(1)}(\bm{r},t+\Delta t)]/2$.
In the second step, the obtained $\hat{h}^{(1)}(\bm{r},t+\Delta t/2)$ is used to update the wavefunctions
\begin{equation}
  \tilde{\psi}_k^{(2)}(\bm{r},t+\Delta t) = \exp\left[-i\hat{h}^{(1)}(\bm{r},t+\Delta t/2)\Delta t\right]\psi_k(\bm{r},t),
\end{equation}
which provide a new estimation for the densities and currents $\tilde{\rho}^{(2)}(t+\Delta t)$ at time $t+\Delta t$.
The Hamiltonian $\hat{h}(\bm{r},t+\Delta t/2)$ in Eq.~\eqref{Eq_pc_solution} is then constructed from the average densities and currents $[\rho(\bm{r},t)+\tilde{\rho}^{(2)}(\bm{r},t+\Delta t)]/2$.

The exponential function of the Hamiltonian operator is evaluated by the Taylor expansion up to order $m$,
\begin{equation}\label{Eq_taylor}
  \exp\left(-i\hat{h}\Delta t\right)\psi\approx\sum_{n = 0}^m\frac{(-i\Delta t)^n}{n!}\hat{h}^n\psi.
\end{equation}
The values of $\Delta t=0.1$~fm/$c$ and $m=4$ are adopted in the following calculations if not specified.
A truncation of the Taylor expansion would violate the strict unitarity of $\exp(-i\hat{h}\Delta t)$ and energy conservation, so the conservation of particle number and energy should be checked carefully to preserve the quality of the time evolution.

\section{Numerical tests}\label{sec_numericaltest}
In this section, the TDCDFT benchmark calculations for the ${}^{16}{\rm O}+{}^{16}{\rm O}$ reaction are performed in 3D lattice space.
Numerical tests, including the excitation energy as a function of boost velocity, the conservation of momentum, total energy, and particle number, as well as  the time reversal invariance, are carefully examined.

\begin{figure}[!htbp]
  \centering
  \includegraphics[width=0.4\textwidth]{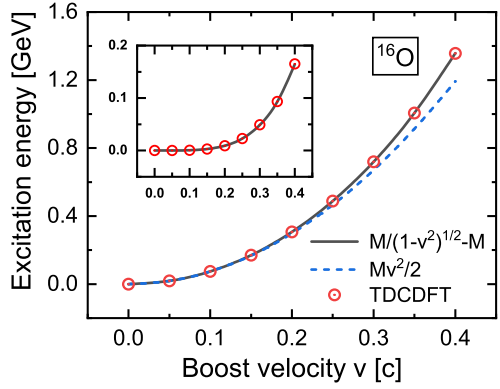}\\
  \caption{(Color online) The excitation energy of a boosted ${}^{16}{\rm O}$ as a function of the boost velocity $v$.
  The open circles represent the excitation energies obtained by TDCDFT.
  The solid and dashed lines denote the results of relativistic $M/(1-v^2)^{1/2}-M$ and nonrelativistic kinetic $Mv^2/2$ energies (see text for the mass $M$), respectively.
  The insert figure shows the results with subtracting the nonrelativistic kinetic energies.
  }\label{fig1}
\end{figure}
The examinations are first focused on the tests involving a single ${}^{16}{\rm O}$.
In Fig.~\ref{fig1}, the excitation energy of a boosted ${}^{16}{\rm O}$ is shown as a function of the boost velocity $v$, whose direction is set along the $z$ axis.
For comparison, the results of relativistic and nonrelativistic kinetic energies, i.e., $M/(1-v^2)^{1/2}-M$ and $Mv^2/2$, are also shown, where the mass $M$ of ${}^{16}{\rm O}$ is evaluated from the ground-state total energy $E_{\rm tot}$ in Eq.~\eqref{Eq_energy_functional}.
The TDCDFT results coincide with the relativistic kinetic energies very well, which is seen more clearly by subtracting the nonrelativistic kinetic energies (see the insert figure in Fig.~\ref{fig1}).
This shows that the adiabatic approximation for $\hat{h}(\bm{r},t)$ in Eq.~\eqref{Eq_td_Dirac_eq} is quite reasonable.
The nonrelativistic kinetic energies deviate from relativistic ones dramatically with the velocity above $0.3c$.

\begin{figure}[!htbp]
  \centering
  \includegraphics[width=0.4\textwidth]{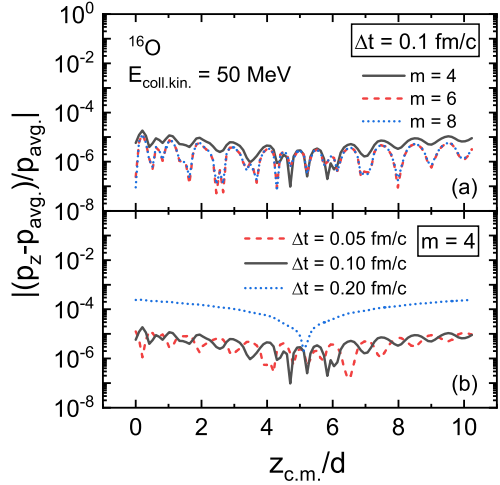}\\
  \caption{(Color online) The relative momentum deviation $|(p_z(t)-p_{\rm avg.})/p_{\rm avg.}|$ with respect to the average momentum $p_{\rm avg.}$ of a boosted ${}^{16}{\rm O}$ as a function of the center-of-mass position $z_{\rm c.m.}$.
  The abscissa is scaled by the mesh size $d$.
  The collective kinetic energy $E_{\rm coll.\,kin.}$ for the boosted ${}^{16}{\rm O}$ is set to $50$ MeV.
  Panel (a) shows the results with the Taylor expansion orders $m = 4$, $6$, $8$ and the time evolution step $\Delta t = 0.10$ fm/$c$.
  Panel (b) shows the results with $\Delta t = 0.05$, $0.10$, $0.20$ fm/$c$ and $m = 4$.
  }\label{fig2}
\end{figure}

A boosted ${}^{16}{\rm O}$ moves with a constant momentum.
In TDCDFT, the momentum $\bm{p}(t)$ is represented by the expectation value of the momentum operator $\hat{\bm{p}}$.
To examine the conservation of momentum, the ${}^{16}{\rm O}$ is placed in the origin point and, then, is boosted with a collective kinetic energy $E_{\rm coll.\,kin.}=50$ MeV along the $z$ axis.
The system is evolved for $T = 100$ fm/$c$.
The average momentum along the $z$ axis is estimated as
\begin{equation}
   p_{\rm avg.}=\frac{\int_0^{T}dt~p_z(t)}{\int_0^{T}dt}.
\end{equation}
Figure \ref{fig2} shows the evolution of the relative momentum deviation $|(p_z(t)-p_{\rm avg.})/p_{\rm avg.}|$ with Taylor expansion orders $m$ and time evolution steps $\Delta t$ as a function of the center-of-mass position $z_{\rm c.m.}$,
which is evaluated by
\begin{equation}
   z_{\rm c.m.}=\frac{\int d^3r~z\rho_v(\bm{r},t)}{\int d^3r~\rho_v(\bm{r},t)}.
\end{equation}

The relative momentum deviation is reduced with larger  $m$ and smaller $\Delta t$.
In the case of $\Delta t = 0.1$ fm/$c$ and $m = 4$, the relative momentum deviations are as small as $10^{-5}$, which reveals the accuracy of the momentum conservation.
Even so, it is interesting to note that the relative momentum deviations oscillate with $z_{\rm c.m.}$, because the space is not exactly translational invariant but is discretized on the lattices.
In fact, the oscillation period is approximately the mesh size $d$.

\begin{figure}[!htbp]
  \centering
  \includegraphics[width=0.4\textwidth]{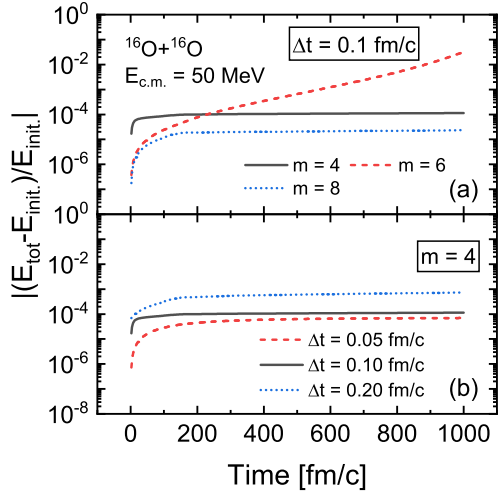}\\
  \caption{(Color online) The relative energy deviation $|(E_{\rm tot}(t)-E_{\rm init.})/E_{\rm init.}|$ with respect to the initial energy $E_{\rm init.}$ for the ${}^{16}{\rm O}+{}^{16}{\rm O}$ head-on collision at the center-of-mass energy $E_{\rm c.m.} = 50$ MeV.
  The rest mass $m_N$ for nucleons has been subtracted from the total energy $E_{\rm tot}$.
  Panel (a) shows the results with the Taylor expansion orders $m = 4$, $6$, $8$ and the time evolution step $\Delta t = 0.10$ fm/$c$.
  Panel (b) shows the results with $\Delta t = 0.05$, $0.10$, $0.20$ fm/$c$ and $m = 4$.
  }\label{fig3}
\end{figure}

Next, the conservation of total energy and particle number, as well as the time reversal invariance for the ${}^{16}{\rm O}+{}^{16}{\rm O}$ reaction are investigated.
The head-on collision with a center-of-mass energy $E_{\rm c.m.}=50$ MeV is taken as an example.

In Fig.~\ref{fig3}, the time evolutions of the relative energy deviation $|(E_{\rm tot}(t)-E_{\rm init.})/E_{\rm init.}|$ with different $\Delta t$ and $m$ values are shown.
For $\Delta t=0.1$ fm/$c$, the relative energy deviations are around $10^{-4}$ and $10^{-5}$ for $m=4$ and $8$, respectively.
However, the evolution of the relative energy deviation for $m=6$ are not stable, in particular at longer time.
The reason is not clear at the moment, but similar phenomenon is also found in the calculation of nonrelativistic TDDFT~\cite{Maruhn2014CPC}.
Moreover, it is found that this unstable behavior for $m=6$ disappears in the calculaions with a smaller $\Delta t$, such as $\Delta t=0.05$ fm/$c$.
For $m=4$, the smaller the time evolution step $\Delta t$, the better the total energy is conserved.
This can be understood because the approximations in Eqs.~\eqref{Eq_pc_solution} and \eqref{Eq_taylor} are better for smaller $\Delta t$ values.

\begin{figure}[!htbp]
  \centering
  \includegraphics[width=0.4\textwidth]{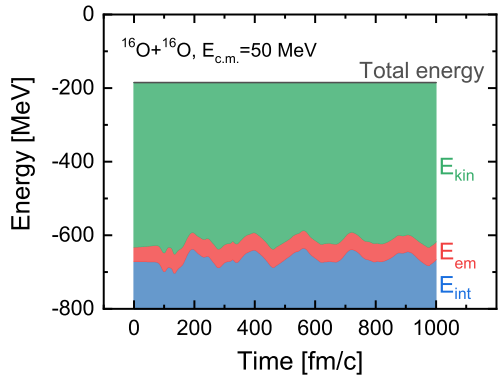}\\
  \caption{(Color online) Time evolution of the total energy and its constituents including the interaction energy $E_{\rm int}$, the electromagnetic energy $E_{\rm em}$, and the kinetic energy $E_{\rm kin}$, for the ${}^{16}{\rm O}+{}^{16}{\rm O}$ head-on collision at the center-of-mass energy $E_{\rm c.m.}=50$ MeV.
  The rest mass $m_N$ for nucleons has been subtracted from the total and kinetic energies.
  }\label{fig4}
\end{figure}

In Fig.~\ref{fig4}, the evolution of the total energy is shown as a function of time, where $\Delta t = 0.1$ fm/$c$ and $m = 4$ are adopted.
The total energy is conserved along the time evolution at a precision about $10^{-4}$.
The three energy constituents including the interaction energy $E_{\rm int}$, the electromagnetic energy $E_{\rm em}$, and the kinetic energy $E_{\rm kin}$ [see Eq.~\eqref{Eq_energy_functional}], are also shown in Fig.~\ref{fig3}.
There are obvious fluctuations up to 70 MeV for these energy constituents, in particular for the interaction and kinetic energies, which correspond to the oscillation of the compound system.
Note that in the present covariant framework, the interaction energy  $E_{\rm int}$ is determined by the densities and/or currents in the scalar and vector channels.
The energy fluctuations in each channel are large and even beyond $1000$ MeV.
This reveals that the conservation of the total energy is indeed achieved by an elegant balance between two large energies in the scalar and vector channels.

\begin{figure}[!htbp]
  \centering
  \includegraphics[width=0.4\textwidth]{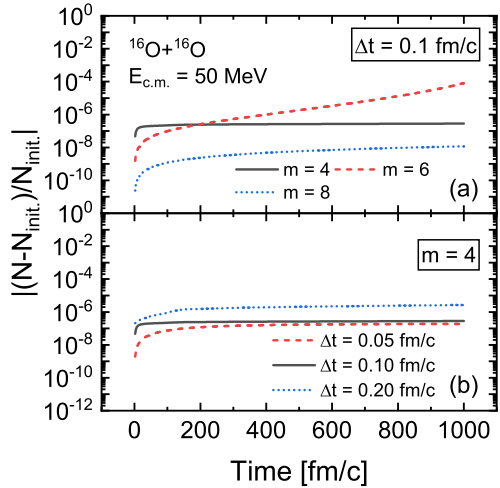}\\
  \caption{(Color online) Same as Fig.~\ref{fig3} but for the relative particle number deviation $|(N(t)-N_{\rm init.})/N_{\rm init.}|$ with respect to the initial particle number $N_{\rm init.}$.
  }\label{fig5}
\end{figure}

Another important examination associated with the approximation in Eq.~\eqref{Eq_taylor} is the conservation of the total particle number $N(t)$ with the definition,
\begin{equation}
  N(t)=\int d^3r~\rho_v(\bm{r},t).
\end{equation}
It reveals the influences of the Taylor expansion on the strict unitarity of the exponential $\exp(-i\hat{h}\Delta t)$.
In Fig.~\ref{fig5}, the time evolution of the relative particle number deviation $|(N(t)-N_{\rm init.})/N_{\rm init.}|$ is shown  with different $\Delta t$ and $m$ values.
Similar to the conservation of the total energy (see Fig.~\ref{fig3}), the particle number is better conserved with smaller $\Delta t$ and larger $m$ values; except for the unstable evolution with $\Delta t=0.1$ fm/$c$ and $m=6$.
The particle number is conserved quite well for all stable evolutions, and the relative particle number deviation is around  $10^{-7}$ at 1000~fm/$c$ in the case of $\Delta t = 0.1$ fm/$c$ and $m = 4$.

All in all, it is found that the momentum, total energy, and particle number are conserved with high precisions in the present TDCDFT calculations with $\Delta t = 0.1$ fm/$c$ and $m = 4$. Therefore, they are adopted in the following investigations.

\begin{figure}[!htbp]
  \centering
  \includegraphics[width=0.4\textwidth]{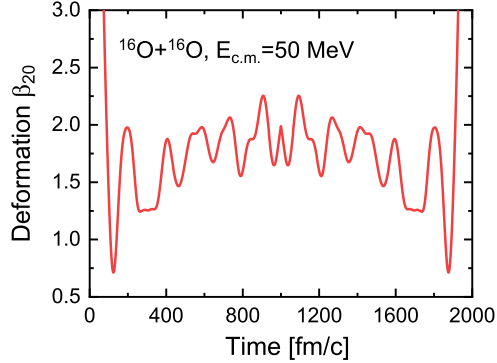}\\
  \caption{(Color online) Time evolution of the quadrupole deformation $\beta_{20}$ for the ${}^{16}{\rm O}$$+^{16}$O head-on collision at $E_{\rm c.m.}=50$ MeV.
  The single-particle wavefunctions at time $t=1000$ fm/$c$ are replaced by their time-reversal conjugates.
  }\label{fig6}
\end{figure}

Apart from the conservation laws, another severe test of the TDCDFT is provided by the time-reversal invariance, which means that the whole system has the microscopic reversibility~\cite{Bonche19761DTDHF, ring2004nuclear}.
To see this property in ${}^{16}{\rm O}$$+^{16}$O head-on collision at $E_{\rm c.m.}=50$ MeV, the single-particle wavefunctions $\psi_k(\bm{r},t)$ at $t=1000$ fm/$c$ are replaced by their time-reversal conjugates,
\begin{equation}
  \hat{T}\psi_k(\bm{r},t)=-i\alpha_x\alpha_z\psi^*_k(\bm{r},t),
\end{equation}
where $\alpha_x$ and $\alpha_z$ are Dirac matrices.
With the time going on, the system should return to the state at the initial time.
In Fig.~\ref{fig6}, the time evolution of the quadrupole deformation $\beta_{20}$ is shown.
It is clearly seen that $\beta_{20}$ evolves back precisely after replacing $\psi_k(\bm{r},t)$ with $\hat{T}\psi_k(\bm{r},t)$ at $1000$~fm/$c$.
Moreover, the nucleon density at $t=2000$ fm/$c$ is also found to agree quite well with the initial one.
These results demonstrate that the time-reversal invariance is fulfilled in the present TDCDFT calculations.

\section{Dissipation dynamics}\label{sec_DisDy}
The dissipation dynamics plays an important role in heavy-ion collisions.
It is responsible for the irreversible conversion of the initial collective kinetic energy into intrinsic nuclear excitations.
To study the dissipation dynamics in deep-inelastic collisions, the ${}^{16}{\rm O}+{}^{16}{\rm O}$ head-on collisions with the center-of-mass energy $E_{\rm c.m.}$ above the upper threshold of fusion are calculated.
A measure of the dissipation is given by the percentage of energy dissipation $P_{\rm dis}=1-E_{\rm fin}/E_{\rm c.m.}$, where $E_{\rm c.m.}$ and $E_{\rm fin}$ represent the initial and final collective kinetic energies, respectively.

\begin{figure}[!htbp]
  \centering
  \includegraphics[width=0.4\textwidth]{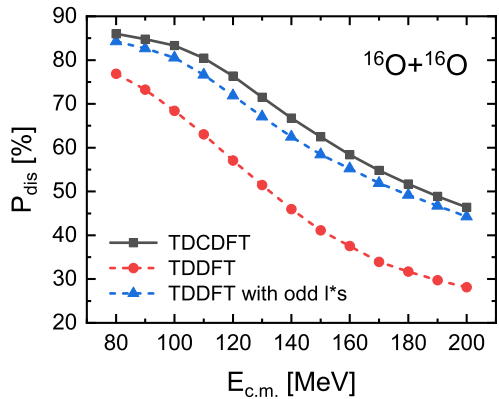}\\
  \caption{(Color online) Percentage of energy dissipation for the ${}^{16}{\rm O}+{}^{16}{\rm O}$ head-on collisions as a function of the center-of-mass energy $E_{\rm c.m.}$.
  For comparison, the nonrelativistic TDDFT results (circle) and the ones with further including the time-odd spin-orbit terms (triangle), taken from Ref.~\cite{Dai2014Dissipation}, are also shown.
  }\label{fig7}
\end{figure}

In Fig.~\ref{fig7}, the percentage of energy dissipation $P_{\rm dis}$ calculated with the TDCDFT is depicted as a function of $E_{\rm c.m.}$ in comparison with the nonrelativistic TDDFT results, which are taken from Ref.~\cite{Dai2014Dissipation}.
The spin-orbit interaction has significant effects on the dissipation, since it couples the spatial motion of the nucleons with the spin degree of freedom, and gives a mechanism for the collective kinetic energy to excite the internal spin degrees of freedom~\cite{STEVENSON2019PPNP}.
It is well-known that the spin-orbit interaction is from relativistic dynamics, and it is naturally taken into account in a covariant density functional.
One can see from Fig.~\ref{fig7} that the energy dissipations $P_{\rm dis}$ in nonrelativistic TDDFT are much lower than the relativistic ones.
The discrepancies are significantly reduced with further including the time-odd spin-orbit terms in the nonrelativistic TDDFT calculations.
This reveals the fact that a covariant density functional automatically contains both time-even and time-odd spin-orbit interactions.

\begin{figure}[!htbp]
  \centering
  \includegraphics[width=0.4\textwidth]{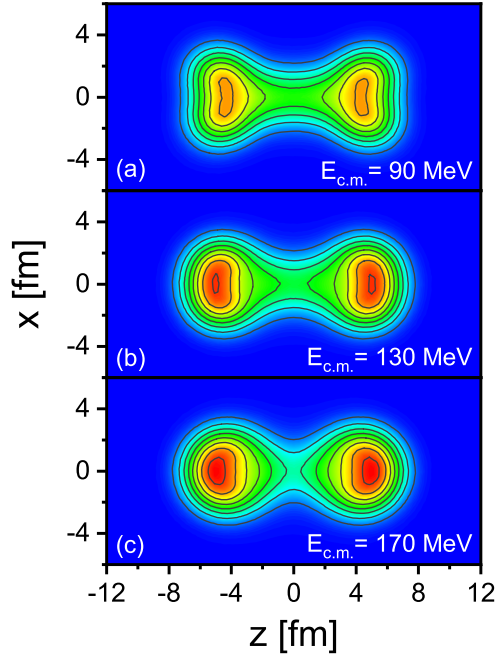}\\
  \caption{(Color online) Density distributions of the separating ions at a given relative distance $R=8.3$ fm for the ${}^{16}{\rm O}+{}^{16}{\rm O}$ head-on collisions with the center-of-mass energies $E_{\rm c.m.}=90$ MeV (top), $130$ MeV (middle), and $170$ MeV (bottom).
  The isolines correspond to multiples of $0.02$~fm$^{-3}$.
  }\label{fig8}
\end{figure}

The features of energy dissipation could be seen more clearly  through the density distributions.
Figure \ref{fig8} shows the density distributions of the separating ions at a given relative distance $R=8.3$ fm for the ${}^{16}{\rm O}+{}^{16}{\rm O}$ head-on collisions with three center-of-mass energies, i.e.,  $E_{\rm c.m.}=90$ MeV, 130 MeV, and 170 MeV.
With the increasing $E_{\rm c.m.}$, the density distribution becomes less diffused.
This is due to the fact that the collective motion becomes faster for larger $E_{\rm c.m.}$ and, thus, the mean field has less time to rearrange itself and more likely keeps its identity as the incident nucleus.
This is also consistent with the decreased trend of the percentage of energy dissipation $P_{\rm dis}$ in Fig.~\ref{fig7}, and for the present three center-of-mass energies, the corresponding $P_{\rm dis}$ is respectively $84.5\%$, $70.9\%$, and $54.2\%$ in the TDCDFT calculations.
Similar features were also obtained in the nonrelativistic TDDFT calculations with the time-odd spin-orbit terms~\cite{Dai2014Dissipation}, while here the density distributions are more diffused in the TDCDFT due to the slightly larger energy dissipation $P_{\rm dis}$ (see Fig.~\ref{fig7}).

\section{Above-barrier fusion cross section }\label{sec_fusion}
The fusion of ${}^{16}{\rm O}+{}^{16}{\rm O}$ at above Coulomb barrier energies is one of the most important benchmarks for the early applications of TDDFT~\cite{KOONIN1976TDHF_O16,Cusson1976TDHFO16, Koonin1977TDHFO16, Flocard1978TDHFO16, Bonche1978TDHFO16, Davies1978TDHFO16}.
The primary reason is that ${}^{16}{\rm O}$ is a light double-magic nucleus, and there are abundant data for the ${}^{16}{\rm O}+{}^{16}{\rm O}$ fusion cross section~\cite{Fernandez1978O16fusion, Tserruya1978O16fusion, Kolata1979O16fusion, Wu1984O16fusion, Thomas1986O16O16fusion}.
The early calculations of TDDFT gave conspicuous transparency for the collisions with low angular momenta, which was, however, not observed in experiment.
This problem is known as the ``fusion window anomaly'', and was latter resolved by the inclusion of spin-orbit interactions~\cite{Umar1986TDHFLS, Reinhard1988TDHFLS}.
Here, the above-barrier fusion cross section of ${}^{16}{\rm O}+{}^{16}{\rm O}$ is investigated with the newly developed TDCDFT in 3D lattice space.

In the present work, the fusion cross section is calculated by
\begin{equation}\label{Eq_fusion}
  \sigma_{\rm fus}(E_{\rm c.m.})=\frac{\pi}{2\mu E_{\rm c.m.}}\sum_{L=0}^{\infty}(2L+1)P_{\rm fus}(L,E_{\rm c.m.}),
\end{equation}
where $\mu$ is the reduced mass of the system, and $P_{\rm fus}(L,E_{\rm c.m.})$ is the fusion probability for the partial wave with orbital angular momentum $L$ at the center-of-mass energy $E_{\rm c.m.}$.
Since ${}^{16}{\rm O}+{}^{16}{\rm O}$ is a system comprised of two identical spin-zero nuclei, the cross section must be multiplied by a factor of 2 and the sum over angular momenta in Eq.~\eqref{Eq_fusion} is restricted to even values of $L$.
Due to the mean-field approximation in TDCDFT, the sub-barrier tunneling of the many-body wavefunction is not included, i.e, $P_{\rm fus}=0$ or $1$.
Such a sharp change can be smoothed by the well-known Hill-Wheeler formula~\cite{Hill1953PhysicalReview} with a Fermi function,
\begin{equation}\label{Eq_HillWheeler}
  P_{\rm fus}(L,E_{\rm c.m.})=\frac{\exp(x_L)}{1+\exp(x_L)},
\end{equation}
with $x_L=[E_{\rm c.m.}-B(L)]/\varepsilon_0$.
Here, the decay constant $\varepsilon_0$ is chosen as $0.4$ MeV~\cite{Esbensen2012HighEdata}, and $B(L)$ is the position of the angular-momentum-dependent barrier.

\begin{figure*}[!htbp]
  \centering
  \includegraphics[width=0.7\textwidth]{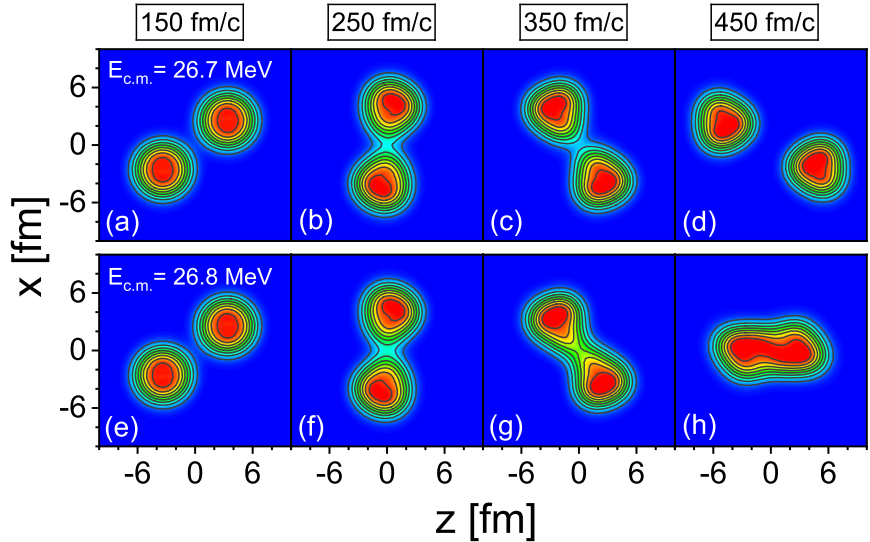}\\
  \caption{(Color online) Total density evolutions for the ${}^{16}{\rm O}+{}^{16}{\rm O}$ reactions with the orbital angular momentum $L=20\hbar$.
  The first and second rows depict the results at the center-of-mass energy $E_{\rm c.m.}=26.7$ MeV and $26.8$ MeV, respectively.
  The isolines correspond to multiples of $0.02$~fm$^{-3}$.
  }\label{fig9}
\end{figure*}

To obtain the barriers $B(L)$ with the TDCDFT, the fusion dynamics are examined in terms of semiclassical trajectories.
As an example, the total density evolutions for the ${}^{16}{\rm O}+{}^{16}{\rm O}$ reactions with $L=20\hbar$ are shown in Fig.~\ref{fig9}.
The first and second rows depict the total density evolutions at the center-of-mass energy $E_{\rm c.m.}=26.7$ MeV and $26.8$ MeV, respectively.
For both energies, the two incident nuclei first form a compound system with a neck [see Figs.~\ref{fig9}(b), (c), (f), and (g)].
The compound system then reseparates in a short time at $E_{\rm c.m.}=26.7$ MeV [see Fig.~\ref{fig9}(d)], while it fuses to a more compact system at $E_{\rm c.m.}=26.8$ MeV [see Fig.~\ref{fig9}(h)].
This indicates that the barrier $B(L=20\hbar)$ is in the range of $26.7\sim26.8$ MeV and, thus, taken as $26.75$ MeV approximately in this work.
The barriers $B(L)$ for other $L$ values can be obtained in the same way, and for a given angular momentum $L$, the center-of-mass energy $E_{\rm c.m.}$ is altered with a step $0.1$ MeV until the transition between not-fusion and fusion is found.

\begin{figure}[!htbp]
  \centering
  \includegraphics[width=0.4\textwidth]{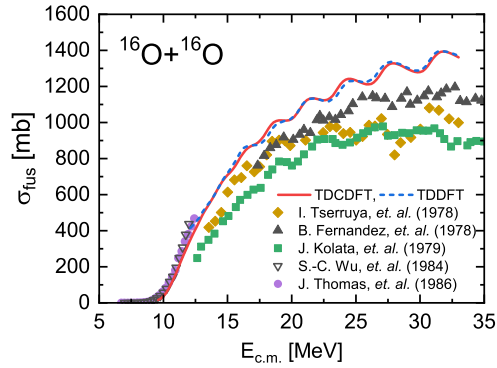}\\
  \caption{(Color online) Above-barrier fusion cross sections as a function of the center-of-mass energy $E_{\rm c.m.}$ for $^{16}{\rm O}+^{16}{\rm O}$ reactions.
The nonrelativistic TDDFT results with the time-odd spin-orbit terms are taken from Ref.~\cite{Simenel2013O16O16}, and the experimental data are taken from Refs.~\cite{Fernandez1978O16fusion, Tserruya1978O16fusion, Kolata1979O16fusion, Wu1984O16fusion, Thomas1986O16O16fusion}.
  }\label{fig10}
\end{figure}

With the obtained barriers $B(L)$, the fusion probability $P_{\rm fus}(L,E_{\rm c.m.})$ can be further calculated via the Hill-Wheeler formula Eq.~\eqref{Eq_HillWheeler}.
The above-barrier fusion cross sections $\sigma_{\rm fus}$ in turn obtained are shown in Fig.~\ref{fig10}, in comparison with the data~\cite{Fernandez1978O16fusion, Tserruya1978O16fusion, Kolata1979O16fusion, Wu1984O16fusion, Thomas1986O16O16fusion} and the nonrelativistic ones.
There is an overall overestimation of the data of Fernandez \textit{et al.}~\cite{Fernandez1978O16fusion} by around $16\%$.
Note that the TDCDFT calculations are based on a universal functional fitted to the bulk properties of the finite nuclei, and have no free parameters coming from the reaction mechanism, so this systematic discrepancy remains small.
Due to the quantization of the angular momentum $L$, the cross sections of the TDCDFT calculations exhibit oscillations with respect to $E_{\rm c.m.}$.
Similar oscillations can also be found in the data.
Therefore, one can conclude that the newly developed TDCDFT in 3D lattice space is an effective approach to investigate the nuclear fusion processes.

For comparison, the nonrelativistic TDDFT results with the time-odd spin-orbit terms~\cite{Simenel2013O16O16} are also shown in Fig.~\ref{fig10}, and they are very close to the TDCDFT ones.
Since the spin-orbit interactions are automatically included in the TDCDFT, here the problem of the fusion window anomaly is resolved naturally; otherwise the fusion cross section would be suppressed significantly~\cite{STEVENSON2019PPNP}.

\section{Summary}\label{sec_summ}
In summary, time-dependent covariant density functional theory with the successful density functional PC-PK1 has been developed in a three-dimensional coordinate space without any symmetry restrictions, and benchmark calculations for the ${}^{16}{\rm O}+{}^{16}{\rm O}$ reaction have been performed systematically.
Numerical tests and two primary applications including the dissipation dynamics and the above-barrier fusion cross sections are performed.
For a boosted ${}^{16}{\rm O}$, the excitation energy with respect to the boost velocity agrees well with the relativistic  kinetic energy, and the total momentum is conserved with a relative deviation around $10^{-5}$ during the time evolution.
For the ${}^{16}{\rm O}+{}^{16}{\rm O}$ head-on collision with the center-of-mass energy $E_{\rm c.m.}=50$ MeV, the total energy and particle number are conserved precisely with the relative deviations respectively around $10^{-4}$ and $10^{-7}$ within a time evolution of 1000 fm/$c$, and the time-reversal invariance is fulfilled quite well.
The dissipation dynamics have been investigated for the deep-inelastic head-on collisions of the  ${}^{16}{\rm O}+{}^{16}{\rm O}$ system.
It is revealed that the obtained percentages of the energy dissipation are reasonable and similar to the nonrelativistic TDDFT results with the time-odd spin-orbit terms.
The above-barrier fusion cross section of ${}^{16}{\rm O}+{}^{16}{\rm O}$ is taken as another benchmark, and the experimental data are well reproduced.
These systematic investigations demonstrate that the TDCDFT in 3D lattice can be an effective approach for the future studies of nuclear dynamical processes.

\begin{acknowledgments}
This work was partly supported by the National Key R\&D Program of China (Contracts No. 2018YFA0404400 and 2017YFE0116700), the National Natural Science Foundation of China (Grants No. 11621131001, 11875075, 11935003, and 11975031), the State Key Laboratory of Nuclear Physics and Technology, Peking University (No. NPT2020ZZ01), and the China Postdoctoral Science Foundation under Grant No. 2020M670013.
\end{acknowledgments}


\begin{thebibliography}{87}%
\makeatletter
\providecommand \@ifxundefined [1]{%
 \@ifx{#1\undefined}
}%
\providecommand \@ifnum [1]{%
 \ifnum #1\expandafter \@firstoftwo
 \else \expandafter \@secondoftwo
 \fi
}%
\providecommand \@ifx [1]{%
 \ifx #1\expandafter \@firstoftwo
 \else \expandafter \@secondoftwo
 \fi
}%
\providecommand \natexlab [1]{#1}%
\providecommand \enquote  [1]{``#1''}%
\providecommand \bibnamefont  [1]{#1}%
\providecommand \bibfnamefont [1]{#1}%
\providecommand \citenamefont [1]{#1}%
\providecommand \href@noop [0]{\@secondoftwo}%
\providecommand \href [0]{\begingroup \@sanitize@url \@href}%
\providecommand \@href[1]{\@@startlink{#1}\@@href}%
\providecommand \@@href[1]{\endgroup#1\@@endlink}%
\providecommand \@sanitize@url [0]{\catcode `\\12\catcode `\$12\catcode
  `\&12\catcode `\#12\catcode `\^12\catcode `\_12\catcode `\%12\relax}%
\providecommand \@@startlink[1]{}%
\providecommand \@@endlink[0]{}%
\providecommand \url  [0]{\begingroup\@sanitize@url \@url }%
\providecommand \@url [1]{\endgroup\@href {#1}{\urlprefix }}%
\providecommand \urlprefix  [0]{URL }%
\providecommand \Eprint [0]{\href }%
\providecommand \doibase [0]{http://dx.doi.org/}%
\providecommand \selectlanguage [0]{\@gobble}%
\providecommand \bibinfo  [0]{\@secondoftwo}%
\providecommand \bibfield  [0]{\@secondoftwo}%
\providecommand \translation [1]{[#1]}%
\providecommand \BibitemOpen [0]{}%
\providecommand \bibitemStop [0]{}%
\providecommand \bibitemNoStop [0]{.\EOS\space}%
\providecommand \EOS [0]{\spacefactor3000\relax}%
\providecommand \BibitemShut  [1]{\csname bibitem#1\endcsname}%
\let\auto@bib@innerbib\@empty
\bibitem [{\citenamefont {Tanihata}\ \emph {et~al.}(2013)\citenamefont
  {Tanihata}, \citenamefont {Savajols},\ and\ \citenamefont
  {Kanungo}}]{Tanihata2013PPNP}%
  \BibitemOpen
  \bibfield  {author} {\bibinfo {author} {\bibfnamefont {I.}~\bibnamefont
  {Tanihata}}, \bibinfo {author} {\bibfnamefont {H.}~\bibnamefont {Savajols}},
  \ and\ \bibinfo {author} {\bibfnamefont {R.}~\bibnamefont {Kanungo}},\ }\href
  {\doibase https://doi.org/10.1016/j.ppnp.2012.07.001} {\bibfield  {journal}
  {\bibinfo  {journal} {Prog. Part. Nucl. Phys.}\ }\textbf {\bibinfo {volume}
  {68}},\ \bibinfo {pages} {215} (\bibinfo {year} {2013})}\BibitemShut
  {NoStop}%
\bibitem [{\citenamefont {Bender}\ \emph {et~al.}(2003)\citenamefont {Bender},
  \citenamefont {Heenen},\ and\ \citenamefont {Reinhard}}]{Bender2003Self}%
  \BibitemOpen
  \bibfield  {author} {\bibinfo {author} {\bibfnamefont {M.}~\bibnamefont
  {Bender}}, \bibinfo {author} {\bibfnamefont {P.-H.}\ \bibnamefont {Heenen}},
  \ and\ \bibinfo {author} {\bibfnamefont {P.-G.}\ \bibnamefont {Reinhard}},\
  }\href {\doibase 10.1103/RevModPhys.75.121} {\bibfield  {journal} {\bibinfo
  {journal} {Rev. Mod. Phys.}\ }\textbf {\bibinfo {volume} {75}},\ \bibinfo
  {pages} {121} (\bibinfo {year} {2003})}\BibitemShut {NoStop}%
\bibitem [{\citenamefont {Meng}(2016)}]{meng2016relativistic}%
  \BibitemOpen
  \bibinfo {editor} {\bibfnamefont {J.}~\bibnamefont {Meng}},\ ed.,\ \href@noop
  {} {\emph {\bibinfo {title} {Relativistic Density Functional for Nuclear
  Structure}}},\ \bibinfo {series} {International Review of Nuclear Physics},
  Vol.~\bibinfo {volume} {10}\ (\bibinfo  {publisher} {World Scientific,
  Singapore},\ \bibinfo {year} {2016})\BibitemShut {NoStop}%
\bibitem [{\citenamefont {Kohn}\ and\ \citenamefont
  {Sham}(1965)}]{Kohn1965DFT}%
  \BibitemOpen
  \bibfield  {author} {\bibinfo {author} {\bibfnamefont {W.}~\bibnamefont
  {Kohn}}\ and\ \bibinfo {author} {\bibfnamefont {L.~J.}\ \bibnamefont
  {Sham}},\ }\href {\doibase 10.1103/PhysRev.140.A1133} {\bibfield  {journal}
  {\bibinfo  {journal} {Phys. Rev.}\ }\textbf {\bibinfo {volume} {140}},\
  \bibinfo {pages} {A1133} (\bibinfo {year} {1965})}\BibitemShut {NoStop}%
\bibitem [{\citenamefont {Ring}(1996)}]{RING1996PPNP}%
  \BibitemOpen
  \bibfield  {author} {\bibinfo {author} {\bibfnamefont {P.}~\bibnamefont
  {Ring}},\ }\href {\doibase http://dx.doi.org/10.1016/0146-6410(96)00054-3}
  {\bibfield  {journal} {\bibinfo  {journal} {Prog. Part. Nucl. Phys.}\
  }\textbf {\bibinfo {volume} {37}},\ \bibinfo {pages} {193} (\bibinfo {year}
  {1996})}\BibitemShut {NoStop}%
\bibitem [{\citenamefont {Vretenar}\ \emph {et~al.}(2005)\citenamefont
  {Vretenar}, \citenamefont {Afanasjev}, \citenamefont {Lalazissis},\ and\
  \citenamefont {Ring}}]{Vretenar2005PhysicsReport}%
  \BibitemOpen
  \bibfield  {author} {\bibinfo {author} {\bibfnamefont {D.}~\bibnamefont
  {Vretenar}}, \bibinfo {author} {\bibfnamefont {A.~V.}\ \bibnamefont
  {Afanasjev}}, \bibinfo {author} {\bibfnamefont {G.~A.}\ \bibnamefont
  {Lalazissis}}, \ and\ \bibinfo {author} {\bibfnamefont {P.}~\bibnamefont
  {Ring}},\ }\href {\doibase http://dx.doi.org/10.1016/j.physrep.2004.10.001}
  {\bibfield  {journal} {\bibinfo  {journal} {Phys. Rep.}\ }\textbf {\bibinfo
  {volume} {409}},\ \bibinfo {pages} {101} (\bibinfo {year}
  {2005})}\BibitemShut {NoStop}%
\bibitem [{\citenamefont {Meng}\ \emph {et~al.}(2006)\citenamefont {Meng},
  \citenamefont {Toki}, \citenamefont {Zhou}, \citenamefont {Zhang},
  \citenamefont {Long},\ and\ \citenamefont {Geng}}]{meng2006PPNP}%
  \BibitemOpen
  \bibfield  {author} {\bibinfo {author} {\bibfnamefont {J.}~\bibnamefont
  {Meng}}, \bibinfo {author} {\bibfnamefont {H.}~\bibnamefont {Toki}}, \bibinfo
  {author} {\bibfnamefont {S.~G.}\ \bibnamefont {Zhou}}, \bibinfo {author}
  {\bibfnamefont {S.~Q.}\ \bibnamefont {Zhang}}, \bibinfo {author}
  {\bibfnamefont {W.~H.}\ \bibnamefont {Long}}, \ and\ \bibinfo {author}
  {\bibfnamefont {L.~S.}\ \bibnamefont {Geng}},\ }\href {\doibase
  https://doi.org/10.1016/j.ppnp.2005.06.001} {\bibfield  {journal} {\bibinfo
  {journal} {Prog. Part. Nucl. Phys.}\ }\textbf {\bibinfo {volume} {57}},\
  \bibinfo {pages} {470} (\bibinfo {year} {2006})}\BibitemShut {NoStop}%
\bibitem [{\citenamefont {Nik\ifmmode \check{s}\else
  \v{s}\fi{}i\ifmmode~\acute{c}\else \'{c}\fi{}}\ \emph
  {et~al.}(2011)\citenamefont {Nik\ifmmode \check{s}\else
  \v{s}\fi{}i\ifmmode~\acute{c}\else \'{c}\fi{}}, \citenamefont {Vretenar},\
  and\ \citenamefont {Ring}}]{NIKSIC2011PPNP}%
  \BibitemOpen
  \bibfield  {author} {\bibinfo {author} {\bibfnamefont {T.}~\bibnamefont
  {Nik\ifmmode \check{s}\else \v{s}\fi{}i\ifmmode~\acute{c}\else \'{c}\fi{}}},
  \bibinfo {author} {\bibfnamefont {D.}~\bibnamefont {Vretenar}}, \ and\
  \bibinfo {author} {\bibfnamefont {P.}~\bibnamefont {Ring}},\ }\href {\doibase
  https://doi.org/10.1016/j.ppnp.2011.01.055} {\bibfield  {journal} {\bibinfo
  {journal} {Prog. Part. Nucl. Phys.}\ }\textbf {\bibinfo {volume} {66}},\
  \bibinfo {pages} {519} (\bibinfo {year} {2011})}\BibitemShut {NoStop}%
\bibitem [{\citenamefont {Serot}\ and\ \citenamefont
  {Walecka}(1986)}]{Volum16}%
  \BibitemOpen
  \bibfield  {author} {\bibinfo {author} {\bibfnamefont {B.~D.}\ \bibnamefont
  {Serot}}\ and\ \bibinfo {author} {\bibfnamefont {J.~D.}\ \bibnamefont
  {Walecka}},\ }\href@noop {} {\bibfield  {journal} {\bibinfo  {journal} {Adv.
  Nucl. Phys.}\ }\textbf {\bibinfo {volume} {16}} (\bibinfo {year}
  {1986})}\BibitemShut {NoStop}%
\bibitem [{\citenamefont {Walecka}(1974)}]{walecka1974theory}%
  \BibitemOpen
  \bibfield  {author} {\bibinfo {author} {\bibfnamefont {J.~D.}\ \bibnamefont
  {Walecka}},\ }\href@noop {} {\bibfield  {journal} {\bibinfo  {journal} {Ann.
  Phys. (NY)}\ }\textbf {\bibinfo {volume} {83}},\ \bibinfo {pages} {491}
  (\bibinfo {year} {1974})}\BibitemShut {NoStop}%
\bibitem [{\citenamefont {Sharma}\ \emph {et~al.}(1995)\citenamefont {Sharma},
  \citenamefont {Lalazissis}, \citenamefont {K\"onig},\ and\ \citenamefont
  {Ring}}]{Sharma1995Pb_shift}%
  \BibitemOpen
  \bibfield  {author} {\bibinfo {author} {\bibfnamefont {M.~M.}\ \bibnamefont
  {Sharma}}, \bibinfo {author} {\bibfnamefont {G.}~\bibnamefont {Lalazissis}},
  \bibinfo {author} {\bibfnamefont {J.}~\bibnamefont {K\"onig}}, \ and\
  \bibinfo {author} {\bibfnamefont {P.}~\bibnamefont {Ring}},\ }\href {\doibase
  10.1103/PhysRevLett.74.3744} {\bibfield  {journal} {\bibinfo  {journal}
  {Phys. Rev. Lett.}\ }\textbf {\bibinfo {volume} {74}},\ \bibinfo {pages}
  {3744} (\bibinfo {year} {1995})}\BibitemShut {NoStop}%
\bibitem [{\citenamefont {Liang}\ \emph {et~al.}(2015)\citenamefont {Liang},
  \citenamefont {Meng},\ and\ \citenamefont {Zhou}}]{liang2015hidden}%
  \BibitemOpen
  \bibfield  {author} {\bibinfo {author} {\bibfnamefont {H.}~\bibnamefont
  {Liang}}, \bibinfo {author} {\bibfnamefont {J.}~\bibnamefont {Meng}}, \ and\
  \bibinfo {author} {\bibfnamefont {S.-G.}\ \bibnamefont {Zhou}},\ }\href
  {\doibase http://dx.doi.org/10.1016/j.physrep.2014.12.005} {\bibfield
  {journal} {\bibinfo  {journal} {Phys. Rep.}\ }\textbf {\bibinfo {volume}
  {570}},\ \bibinfo {pages} {1} (\bibinfo {year} {2015})}\BibitemShut {NoStop}%
\bibitem [{\citenamefont {Meng}\ \emph {et~al.}(2013)\citenamefont {Meng},
  \citenamefont {Peng}, \citenamefont {Zhang},\ and\ \citenamefont
  {Zhao}}]{Meng2013FT_TAC}%
  \BibitemOpen
  \bibfield  {author} {\bibinfo {author} {\bibfnamefont {J.}~\bibnamefont
  {Meng}}, \bibinfo {author} {\bibfnamefont {J.}~\bibnamefont {Peng}}, \bibinfo
  {author} {\bibfnamefont {S.-Q.}\ \bibnamefont {Zhang}}, \ and\ \bibinfo
  {author} {\bibfnamefont {P.-W.}\ \bibnamefont {Zhao}},\ }\href {\doibase
  10.1007/s11467-013-0287-y} {\bibfield  {journal} {\bibinfo  {journal} {Front.
  Phys.}\ }\textbf {\bibinfo {volume} {8}},\ \bibinfo {pages} {55} (\bibinfo
  {year} {2013})}\BibitemShut {NoStop}%
\bibitem [{\citenamefont {Meng}\ and\ \citenamefont
  {Ring}(1996)}]{meng1996relativistic}%
  \BibitemOpen
  \bibfield  {author} {\bibinfo {author} {\bibfnamefont {J.}~\bibnamefont
  {Meng}}\ and\ \bibinfo {author} {\bibfnamefont {P.}~\bibnamefont {Ring}},\
  }\href {\doibase 10.1103/PhysRevLett.77.3963} {\bibfield  {journal} {\bibinfo
   {journal} {Phys. Rev. Lett.}\ }\textbf {\bibinfo {volume} {77}},\ \bibinfo
  {pages} {3963} (\bibinfo {year} {1996})}\BibitemShut {NoStop}%
\bibitem [{\citenamefont {Meng}\ and\ \citenamefont
  {Ring}(1998)}]{meng1998giant}%
  \BibitemOpen
  \bibfield  {author} {\bibinfo {author} {\bibfnamefont {J.}~\bibnamefont
  {Meng}}\ and\ \bibinfo {author} {\bibfnamefont {P.}~\bibnamefont {Ring}},\
  }\href {\doibase 10.1103/PhysRevLett.80.460} {\bibfield  {journal} {\bibinfo
  {journal} {Phys. Rev. Lett.}\ }\textbf {\bibinfo {volume} {80}},\ \bibinfo
  {pages} {460} (\bibinfo {year} {1998})}\BibitemShut {NoStop}%
\bibitem [{\citenamefont {Zhou}\ \emph {et~al.}(2010)\citenamefont {Zhou},
  \citenamefont {Meng}, \citenamefont {Ring},\ and\ \citenamefont
  {Zhao}}]{zhou2010neutron}%
  \BibitemOpen
  \bibfield  {author} {\bibinfo {author} {\bibfnamefont {S.-G.}\ \bibnamefont
  {Zhou}}, \bibinfo {author} {\bibfnamefont {J.}~\bibnamefont {Meng}}, \bibinfo
  {author} {\bibfnamefont {P.}~\bibnamefont {Ring}}, \ and\ \bibinfo {author}
  {\bibfnamefont {E.-G.}\ \bibnamefont {Zhao}},\ }\href {\doibase
  10.1103/PhysRevC.82.011301} {\bibfield  {journal} {\bibinfo  {journal} {Phys.
  Rev. C}\ }\textbf {\bibinfo {volume} {82}},\ \bibinfo {pages} {011301}
  (\bibinfo {year} {2010})}\BibitemShut {NoStop}%
\bibitem [{\citenamefont {Xia}\ \emph {et~al.}(2018)\citenamefont {Xia},
  \citenamefont {Lim}, \citenamefont {Zhao}, \citenamefont {Liang},
  \citenamefont {Qu}, \citenamefont {Chen}, \citenamefont {Liu}, \citenamefont
  {Zhang}, \citenamefont {Zhang}, \citenamefont {Kim},\ and\ \citenamefont
  {Meng}}]{xia2018ADNDT}%
  \BibitemOpen
  \bibfield  {author} {\bibinfo {author} {\bibfnamefont {X.~W.}\ \bibnamefont
  {Xia}}, \bibinfo {author} {\bibfnamefont {Y.}~\bibnamefont {Lim}}, \bibinfo
  {author} {\bibfnamefont {P.~W.}\ \bibnamefont {Zhao}}, \bibinfo {author}
  {\bibfnamefont {H.~Z.}\ \bibnamefont {Liang}}, \bibinfo {author}
  {\bibfnamefont {X.~Y.}\ \bibnamefont {Qu}}, \bibinfo {author} {\bibfnamefont
  {Y.}~\bibnamefont {Chen}}, \bibinfo {author} {\bibfnamefont {H.}~\bibnamefont
  {Liu}}, \bibinfo {author} {\bibfnamefont {L.~F.}\ \bibnamefont {Zhang}},
  \bibinfo {author} {\bibfnamefont {S.~Q.}\ \bibnamefont {Zhang}}, \bibinfo
  {author} {\bibfnamefont {Y.}~\bibnamefont {Kim}}, \ and\ \bibinfo {author}
  {\bibfnamefont {J.}~\bibnamefont {Meng}},\ }\href {\doibase
  https://doi.org/10.1016/j.adt.2017.09.001} {\bibfield  {journal} {\bibinfo
  {journal} {At. Data Nucl. Data Tables}\ }\textbf {\bibinfo {volume}
  {121-122}},\ \bibinfo {pages} {1} (\bibinfo {year} {2018})}\BibitemShut
  {NoStop}%
\bibitem [{\citenamefont {Peng}\ \emph {et~al.}(2008)\citenamefont {Peng},
  \citenamefont {Meng}, \citenamefont {Ring},\ and\ \citenamefont
  {Zhang}}]{Peng2008maganetic_roration}%
  \BibitemOpen
  \bibfield  {author} {\bibinfo {author} {\bibfnamefont {J.}~\bibnamefont
  {Peng}}, \bibinfo {author} {\bibfnamefont {J.}~\bibnamefont {Meng}}, \bibinfo
  {author} {\bibfnamefont {P.}~\bibnamefont {Ring}}, \ and\ \bibinfo {author}
  {\bibfnamefont {S.~Q.}\ \bibnamefont {Zhang}},\ }\href {\doibase
  10.1103/PhysRevC.78.024313} {\bibfield  {journal} {\bibinfo  {journal} {Phys.
  Rev. C}\ }\textbf {\bibinfo {volume} {78}},\ \bibinfo {pages} {024313}
  (\bibinfo {year} {2008})}\BibitemShut {NoStop}%
\bibitem [{\citenamefont {Zhao}\ \emph {et~al.}(2011)\citenamefont {Zhao},
  \citenamefont {Peng}, \citenamefont {Liang}, \citenamefont {Ring},\ and\
  \citenamefont {Meng}}]{Zhao2011PRL_AMR}%
  \BibitemOpen
  \bibfield  {author} {\bibinfo {author} {\bibfnamefont {P.~W.}\ \bibnamefont
  {Zhao}}, \bibinfo {author} {\bibfnamefont {J.}~\bibnamefont {Peng}}, \bibinfo
  {author} {\bibfnamefont {H.~Z.}\ \bibnamefont {Liang}}, \bibinfo {author}
  {\bibfnamefont {P.}~\bibnamefont {Ring}}, \ and\ \bibinfo {author}
  {\bibfnamefont {J.}~\bibnamefont {Meng}},\ }\href {\doibase
  10.1103/PhysRevLett.107.122501} {\bibfield  {journal} {\bibinfo  {journal}
  {Phys. Rev. Lett.}\ }\textbf {\bibinfo {volume} {107}},\ \bibinfo {pages}
  {122501} (\bibinfo {year} {2011})}\BibitemShut {NoStop}%
\bibitem [{\citenamefont {Zhao}\ \emph {et~al.}(2015)\citenamefont {Zhao},
  \citenamefont {Itagaki},\ and\ \citenamefont {Meng}}]{Zhao2015Rod-shaped}%
  \BibitemOpen
  \bibfield  {author} {\bibinfo {author} {\bibfnamefont {P.~W.}\ \bibnamefont
  {Zhao}}, \bibinfo {author} {\bibfnamefont {N.}~\bibnamefont {Itagaki}}, \
  and\ \bibinfo {author} {\bibfnamefont {J.}~\bibnamefont {Meng}},\ }\href
  {\doibase 10.1103/PhysRevLett.115.022501} {\bibfield  {journal} {\bibinfo
  {journal} {Phys. Rev. Lett.}\ }\textbf {\bibinfo {volume} {115}},\ \bibinfo
  {pages} {022501} (\bibinfo {year} {2015})}\BibitemShut {NoStop}%
\bibitem [{\citenamefont {Zhao}(2017)}]{Zhao2017ChiralRotation}%
  \BibitemOpen
  \bibfield  {author} {\bibinfo {author} {\bibfnamefont {P.~W.}\ \bibnamefont
  {Zhao}},\ }\href
  {http://www.sciencedirect.com/science/article/pii/S0370269317306226}
  {\bibfield  {journal} {\bibinfo  {journal} {Phys. Lett. B}\ }\textbf
  {\bibinfo {volume} {773}},\ \bibinfo {pages} {1} (\bibinfo {year}
  {2017})}\BibitemShut {NoStop}%
\bibitem [{\citenamefont {Nik\ifmmode \check{s}\else
  \v{s}\fi{}i\ifmmode~\acute{c}\else \'{c}\fi{}}\ \emph
  {et~al.}(2002)\citenamefont {Nik\ifmmode \check{s}\else
  \v{s}\fi{}i\ifmmode~\acute{c}\else \'{c}\fi{}}, \citenamefont {Vretenar},\
  and\ \citenamefont {Ring}}]{Niksic2002DDME1_QRPA}%
  \BibitemOpen
  \bibfield  {author} {\bibinfo {author} {\bibfnamefont {T.}~\bibnamefont
  {Nik\ifmmode \check{s}\else \v{s}\fi{}i\ifmmode~\acute{c}\else \'{c}\fi{}}},
  \bibinfo {author} {\bibfnamefont {D.}~\bibnamefont {Vretenar}}, \ and\
  \bibinfo {author} {\bibfnamefont {P.}~\bibnamefont {Ring}},\ }\href {\doibase
  10.1103/PhysRevC.66.064302} {\bibfield  {journal} {\bibinfo  {journal} {Phys.
  Rev. C}\ }\textbf {\bibinfo {volume} {66}},\ \bibinfo {pages} {064302}
  (\bibinfo {year} {2002})}\BibitemShut {NoStop}%
\bibitem [{\citenamefont {Paar}\ \emph {et~al.}(2007)\citenamefont {Paar},
  \citenamefont {Vretenar}, \citenamefont {Khan},\ and\ \citenamefont
  {Col{\`{o}}}}]{Paar2007RPP}%
  \BibitemOpen
  \bibfield  {author} {\bibinfo {author} {\bibfnamefont {N.}~\bibnamefont
  {Paar}}, \bibinfo {author} {\bibfnamefont {D.}~\bibnamefont {Vretenar}},
  \bibinfo {author} {\bibfnamefont {E.}~\bibnamefont {Khan}}, \ and\ \bibinfo
  {author} {\bibfnamefont {G.}~\bibnamefont {Col{\`{o}}}},\ }\href {\doibase
  10.1088/0034-4885/70/5/r02} {\bibfield  {journal} {\bibinfo  {journal} {Rep.
  Prog. Phys.}\ }\textbf {\bibinfo {volume} {70}},\ \bibinfo {pages} {691}
  (\bibinfo {year} {2007})}\BibitemShut {NoStop}%
\bibitem [{\citenamefont {Paar}\ \emph {et~al.}(2009)\citenamefont {Paar},
  \citenamefont {Niu}, \citenamefont {Vretenar},\ and\ \citenamefont
  {Meng}}]{Paar2009QRPA}%
  \BibitemOpen
  \bibfield  {author} {\bibinfo {author} {\bibfnamefont {N.}~\bibnamefont
  {Paar}}, \bibinfo {author} {\bibfnamefont {Y.~F.}\ \bibnamefont {Niu}},
  \bibinfo {author} {\bibfnamefont {D.}~\bibnamefont {Vretenar}}, \ and\
  \bibinfo {author} {\bibfnamefont {J.}~\bibnamefont {Meng}},\ }\href {\doibase
  10.1103/PhysRevLett.103.032502} {\bibfield  {journal} {\bibinfo  {journal}
  {Phys. Rev. Lett.}\ }\textbf {\bibinfo {volume} {103}},\ \bibinfo {pages}
  {032502} (\bibinfo {year} {2009})}\BibitemShut {NoStop}%
\bibitem [{\citenamefont {Niu}\ \emph {et~al.}(2009)\citenamefont {Niu},
  \citenamefont {Paar}, \citenamefont {Vretenar},\ and\ \citenamefont
  {Meng}}]{Niu2009FTQRPA}%
  \BibitemOpen
  \bibfield  {author} {\bibinfo {author} {\bibfnamefont {Y.}~\bibnamefont
  {Niu}}, \bibinfo {author} {\bibfnamefont {N.}~\bibnamefont {Paar}}, \bibinfo
  {author} {\bibfnamefont {D.}~\bibnamefont {Vretenar}}, \ and\ \bibinfo
  {author} {\bibfnamefont {J.}~\bibnamefont {Meng}},\ }\href {\doibase
  https://doi.org/10.1016/j.physletb.2009.10.046} {\bibfield  {journal}
  {\bibinfo  {journal} {Phys. Lett. B}\ }\textbf {\bibinfo {volume} {681}},\
  \bibinfo {pages} {315} (\bibinfo {year} {2009})}\BibitemShut {NoStop}%
\bibitem [{\citenamefont {Runge}\ and\ \citenamefont
  {Gross}(1984)}]{Rung1984TDDFT}%
  \BibitemOpen
  \bibfield  {author} {\bibinfo {author} {\bibfnamefont {E.}~\bibnamefont
  {Runge}}\ and\ \bibinfo {author} {\bibfnamefont {E.~K.~U.}\ \bibnamefont
  {Gross}},\ }\href {\doibase 10.1103/PhysRevLett.52.997} {\bibfield  {journal}
  {\bibinfo  {journal} {Phys. Rev. Lett.}\ }\textbf {\bibinfo {volume} {52}},\
  \bibinfo {pages} {997} (\bibinfo {year} {1984})}\BibitemShut {NoStop}%
\bibitem [{\citenamefont {Engel}\ \emph {et~al.}(1975)\citenamefont {Engel},
  \citenamefont {Brink}, \citenamefont {Goeke}, \citenamefont {Krieger},\ and\
  \citenamefont {Vautherin}}]{ENGEL1975215NPA}%
  \BibitemOpen
  \bibfield  {author} {\bibinfo {author} {\bibfnamefont {Y.~M.}\ \bibnamefont
  {Engel}}, \bibinfo {author} {\bibfnamefont {D.~M.}\ \bibnamefont {Brink}},
  \bibinfo {author} {\bibfnamefont {K.}~\bibnamefont {Goeke}}, \bibinfo
  {author} {\bibfnamefont {S.~J.}\ \bibnamefont {Krieger}}, \ and\ \bibinfo
  {author} {\bibfnamefont {D.}~\bibnamefont {Vautherin}},\ }\href {\doibase
  http://dx.doi.org/10.1016/0375-9474(75)90184-0} {\bibfield  {journal}
  {\bibinfo  {journal} {Nucl. Phys. A}\ }\textbf {\bibinfo {volume} {249}},\
  \bibinfo {pages} {215} (\bibinfo {year} {1975})}\BibitemShut {NoStop}%
\bibitem [{\citenamefont {Bonche}\ \emph {et~al.}(1976)\citenamefont {Bonche},
  \citenamefont {Koonin},\ and\ \citenamefont {Negele}}]{Bonche19761DTDHF}%
  \BibitemOpen
  \bibfield  {author} {\bibinfo {author} {\bibfnamefont {P.}~\bibnamefont
  {Bonche}}, \bibinfo {author} {\bibfnamefont {S.}~\bibnamefont {Koonin}}, \
  and\ \bibinfo {author} {\bibfnamefont {J.~W.}\ \bibnamefont {Negele}},\
  }\href {\doibase 10.1103/PhysRevC.13.1226} {\bibfield  {journal} {\bibinfo
  {journal} {Phys. Rev. C}\ }\textbf {\bibinfo {volume} {13}},\ \bibinfo
  {pages} {1226} (\bibinfo {year} {1976})}\BibitemShut {NoStop}%
\bibitem [{\citenamefont {Koonin}(1976)}]{KOONIN1976TDHF_O16}%
  \BibitemOpen
  \bibfield  {author} {\bibinfo {author} {\bibfnamefont {S.}~\bibnamefont
  {Koonin}},\ }\href {\doibase https://doi.org/10.1016/0370-2693(76)90135-0}
  {\bibfield  {journal} {\bibinfo  {journal} {Phys. Lett. B}\ }\textbf
  {\bibinfo {volume} {61}},\ \bibinfo {pages} {227} (\bibinfo {year}
  {1976})}\BibitemShut {NoStop}%
\bibitem [{\citenamefont {Cusson}\ \emph {et~al.}(1976)\citenamefont {Cusson},
  \citenamefont {Smith},\ and\ \citenamefont {Maruhn}}]{Cusson1976TDHFO16}%
  \BibitemOpen
  \bibfield  {author} {\bibinfo {author} {\bibfnamefont {R.~Y.}\ \bibnamefont
  {Cusson}}, \bibinfo {author} {\bibfnamefont {R.~K.}\ \bibnamefont {Smith}}, \
  and\ \bibinfo {author} {\bibfnamefont {J.~A.}\ \bibnamefont {Maruhn}},\
  }\href {\doibase 10.1103/PhysRevLett.36.1166} {\bibfield  {journal} {\bibinfo
   {journal} {Phys. Rev. Lett.}\ }\textbf {\bibinfo {volume} {36}},\ \bibinfo
  {pages} {1166} (\bibinfo {year} {1976})}\BibitemShut {NoStop}%
\bibitem [{\citenamefont {Koonin}\ \emph {et~al.}(1977)\citenamefont {Koonin},
  \citenamefont {Davies}, \citenamefont {Maruhn-Rezwani}, \citenamefont
  {Feldmeier}, \citenamefont {Krieger},\ and\ \citenamefont
  {Negele}}]{Koonin1977TDHFO16}%
  \BibitemOpen
  \bibfield  {author} {\bibinfo {author} {\bibfnamefont {S.~E.}\ \bibnamefont
  {Koonin}}, \bibinfo {author} {\bibfnamefont {K.~T.~R.}\ \bibnamefont
  {Davies}}, \bibinfo {author} {\bibfnamefont {V.}~\bibnamefont
  {Maruhn-Rezwani}}, \bibinfo {author} {\bibfnamefont {H.}~\bibnamefont
  {Feldmeier}}, \bibinfo {author} {\bibfnamefont {S.~J.}\ \bibnamefont
  {Krieger}}, \ and\ \bibinfo {author} {\bibfnamefont {J.~W.}\ \bibnamefont
  {Negele}},\ }\href {\doibase 10.1103/PhysRevC.15.1359} {\bibfield  {journal}
  {\bibinfo  {journal} {Phys. Rev. C}\ }\textbf {\bibinfo {volume} {15}},\
  \bibinfo {pages} {1359} (\bibinfo {year} {1977})}\BibitemShut {NoStop}%
\bibitem [{\citenamefont {Flocard}\ \emph {et~al.}(1978)\citenamefont
  {Flocard}, \citenamefont {Koonin},\ and\ \citenamefont
  {Weiss}}]{Flocard1978TDHFO16}%
  \BibitemOpen
  \bibfield  {author} {\bibinfo {author} {\bibfnamefont {H.}~\bibnamefont
  {Flocard}}, \bibinfo {author} {\bibfnamefont {S.~E.}\ \bibnamefont {Koonin}},
  \ and\ \bibinfo {author} {\bibfnamefont {M.~S.}\ \bibnamefont {Weiss}},\
  }\href {\doibase 10.1103/PhysRevC.17.1682} {\bibfield  {journal} {\bibinfo
  {journal} {Phys. Rev. C}\ }\textbf {\bibinfo {volume} {17}},\ \bibinfo
  {pages} {1682} (\bibinfo {year} {1978})}\BibitemShut {NoStop}%
\bibitem [{\citenamefont {Bonche}\ \emph {et~al.}(1978)\citenamefont {Bonche},
  \citenamefont {Grammaticos},\ and\ \citenamefont
  {Koonin}}]{Bonche1978TDHFO16}%
  \BibitemOpen
  \bibfield  {author} {\bibinfo {author} {\bibfnamefont {P.}~\bibnamefont
  {Bonche}}, \bibinfo {author} {\bibfnamefont {B.}~\bibnamefont {Grammaticos}},
  \ and\ \bibinfo {author} {\bibfnamefont {S.}~\bibnamefont {Koonin}},\ }\href
  {\doibase 10.1103/PhysRevC.17.1700} {\bibfield  {journal} {\bibinfo
  {journal} {Phys. Rev. C}\ }\textbf {\bibinfo {volume} {17}},\ \bibinfo
  {pages} {1700} (\bibinfo {year} {1978})}\BibitemShut {NoStop}%
\bibitem [{\citenamefont {Davies}\ \emph {et~al.}(1978)\citenamefont {Davies},
  \citenamefont {Feldmeier}, \citenamefont {Flocard},\ and\ \citenamefont
  {Weiss}}]{Davies1978TDHFO16}%
  \BibitemOpen
  \bibfield  {author} {\bibinfo {author} {\bibfnamefont {K.~T.~R.}\
  \bibnamefont {Davies}}, \bibinfo {author} {\bibfnamefont {H.~T.}\
  \bibnamefont {Feldmeier}}, \bibinfo {author} {\bibfnamefont {H.}~\bibnamefont
  {Flocard}}, \ and\ \bibinfo {author} {\bibfnamefont {M.~S.}\ \bibnamefont
  {Weiss}},\ }\href {\doibase 10.1103/PhysRevC.18.2631} {\bibfield  {journal}
  {\bibinfo  {journal} {Phys. Rev. C}\ }\textbf {\bibinfo {volume} {18}},\
  \bibinfo {pages} {2631} (\bibinfo {year} {1978})}\BibitemShut {NoStop}%
\bibitem [{\citenamefont {Dirac}(1930)}]{dirac1930TDHF}%
  \BibitemOpen
  \bibfield  {author} {\bibinfo {author} {\bibfnamefont {P.~A.~M.}\
  \bibnamefont {Dirac}},\ }\href {\doibase 10.1017/S0305004100016108}
  {\bibfield  {journal} {\bibinfo  {journal} {Math. Proc. Cambridge}\ }\textbf
  {\bibinfo {volume} {26}},\ \bibinfo {pages} {376} (\bibinfo {year}
  {1930})}\BibitemShut {NoStop}%
\bibitem [{\citenamefont {Negele}(1982)}]{Negele1982TDDFT}%
  \BibitemOpen
  \bibfield  {author} {\bibinfo {author} {\bibfnamefont {J.~W.}\ \bibnamefont
  {Negele}},\ }\href {\doibase 10.1103/RevModPhys.54.913} {\bibfield  {journal}
  {\bibinfo  {journal} {Rev. Mod. Phys.}\ }\textbf {\bibinfo {volume} {54}},\
  \bibinfo {pages} {913} (\bibinfo {year} {1982})}\BibitemShut {NoStop}%
\bibitem [{\citenamefont {Simenel}(2012)}]{Simene2012PEPJA}%
  \BibitemOpen
  \bibfield  {author} {\bibinfo {author} {\bibfnamefont {C.}~\bibnamefont
  {Simenel}},\ }\href {\doibase 10.1140/epja/i2012-12152-0} {\bibfield
  {journal} {\bibinfo  {journal} {Eur. Phys. J. A}\ }\textbf {\bibinfo {volume}
  {48}},\ \bibinfo {pages} {152} (\bibinfo {year} {2012})}\BibitemShut
  {NoStop}%
\bibitem [{\citenamefont {Nakatsukasa}\ \emph {et~al.}(2016)\citenamefont
  {Nakatsukasa}, \citenamefont {Matsuyanagi}, \citenamefont {Matsuo},\ and\
  \citenamefont {Yabana}}]{NakatsukasaRMP2016}%
  \BibitemOpen
  \bibfield  {author} {\bibinfo {author} {\bibfnamefont {T.}~\bibnamefont
  {Nakatsukasa}}, \bibinfo {author} {\bibfnamefont {K.}~\bibnamefont
  {Matsuyanagi}}, \bibinfo {author} {\bibfnamefont {M.}~\bibnamefont {Matsuo}},
  \ and\ \bibinfo {author} {\bibfnamefont {K.}~\bibnamefont {Yabana}},\ }\href
  {\doibase 10.1103/RevModPhys.88.045004} {\bibfield  {journal} {\bibinfo
  {journal} {Rev. Mod. Phys.}\ }\textbf {\bibinfo {volume} {88}},\ \bibinfo
  {pages} {045004} (\bibinfo {year} {2016})}\BibitemShut {NoStop}%
\bibitem [{\citenamefont {Simenel}\ and\ \citenamefont
  {Umar}(2018)}]{SIMENEL2018TDHF_PPNP}%
  \BibitemOpen
  \bibfield  {author} {\bibinfo {author} {\bibfnamefont {C.}~\bibnamefont
  {Simenel}}\ and\ \bibinfo {author} {\bibfnamefont {A.~S.}\ \bibnamefont
  {Umar}},\ }\href {\doibase https://doi.org/10.1016/j.ppnp.2018.07.002}
  {\bibfield  {journal} {\bibinfo  {journal} {Prog. Part. Nucl. Phys.}\
  }\textbf {\bibinfo {volume} {103}},\ \bibinfo {pages} {19} (\bibinfo {year}
  {2018})}\BibitemShut {NoStop}%
\bibitem [{\citenamefont {Stevenson}\ and\ \citenamefont
  {Barton}(2019)}]{STEVENSON2019PPNP}%
  \BibitemOpen
  \bibfield  {author} {\bibinfo {author} {\bibfnamefont {P.~D.}\ \bibnamefont
  {Stevenson}}\ and\ \bibinfo {author} {\bibfnamefont {M.~C.}\ \bibnamefont
  {Barton}},\ }\href {\doibase https://doi.org/10.1016/j.ppnp.2018.09.002}
  {\bibfield  {journal} {\bibinfo  {journal} {Prog. Part. Nucl. Phys.}\
  }\textbf {\bibinfo {volume} {104}},\ \bibinfo {pages} {142} (\bibinfo {year}
  {2019})}\BibitemShut {NoStop}%
\bibitem [{\citenamefont {Simenel}(2010)}]{Simenel2010MNT}%
  \BibitemOpen
  \bibfield  {author} {\bibinfo {author} {\bibfnamefont {C.}~\bibnamefont
  {Simenel}},\ }\href {\doibase 10.1103/PhysRevLett.105.192701} {\bibfield
  {journal} {\bibinfo  {journal} {Phys. Rev. Lett.}\ }\textbf {\bibinfo
  {volume} {105}},\ \bibinfo {pages} {192701} (\bibinfo {year}
  {2010})}\BibitemShut {NoStop}%
\bibitem [{\citenamefont {Sekizawa}\ and\ \citenamefont
  {Yabana}(2013)}]{Sekizawa2013TDHFMNT}%
  \BibitemOpen
  \bibfield  {author} {\bibinfo {author} {\bibfnamefont {K.}~\bibnamefont
  {Sekizawa}}\ and\ \bibinfo {author} {\bibfnamefont {K.}~\bibnamefont
  {Yabana}},\ }\href {\doibase 10.1103/PhysRevC.88.014614} {\bibfield
  {journal} {\bibinfo  {journal} {Phys. Rev. C}\ }\textbf {\bibinfo {volume}
  {88}},\ \bibinfo {pages} {014614} (\bibinfo {year} {2013})}\BibitemShut
  {NoStop}%
\bibitem [{\citenamefont {Sekizawa}\ and\ \citenamefont
  {Yabana}(2016)}]{Sekizawa2016TDHF_Ni_U}%
  \BibitemOpen
  \bibfield  {author} {\bibinfo {author} {\bibfnamefont {K.}~\bibnamefont
  {Sekizawa}}\ and\ \bibinfo {author} {\bibfnamefont {K.}~\bibnamefont
  {Yabana}},\ }\href {\doibase 10.1103/PhysRevC.93.054616} {\bibfield
  {journal} {\bibinfo  {journal} {Phys. Rev. C}\ }\textbf {\bibinfo {volume}
  {93}},\ \bibinfo {pages} {054616} (\bibinfo {year} {2016})}\BibitemShut
  {NoStop}%
\bibitem [{\citenamefont {Wu}\ and\ \citenamefont {Guo}(2019)}]{Wu2019MNT}%
  \BibitemOpen
  \bibfield  {author} {\bibinfo {author} {\bibfnamefont {Z.}~\bibnamefont
  {Wu}}\ and\ \bibinfo {author} {\bibfnamefont {L.}~\bibnamefont {Guo}},\
  }\href {\doibase 10.1103/PhysRevC.100.014612} {\bibfield  {journal} {\bibinfo
   {journal} {Phys. Rev. C}\ }\textbf {\bibinfo {volume} {100}},\ \bibinfo
  {pages} {014612} (\bibinfo {year} {2019})}\BibitemShut {NoStop}%
\bibitem [{\citenamefont {Goddard}\ \emph {et~al.}(2015)\citenamefont
  {Goddard}, \citenamefont {Stevenson},\ and\ \citenamefont
  {Rios}}]{Goddard2015TDHFfission}%
  \BibitemOpen
  \bibfield  {author} {\bibinfo {author} {\bibfnamefont {P.}~\bibnamefont
  {Goddard}}, \bibinfo {author} {\bibfnamefont {P.}~\bibnamefont {Stevenson}},
  \ and\ \bibinfo {author} {\bibfnamefont {A.}~\bibnamefont {Rios}},\ }\href
  {\doibase 10.1103/PhysRevC.92.054610} {\bibfield  {journal} {\bibinfo
  {journal} {Phys. Rev. C}\ }\textbf {\bibinfo {volume} {92}},\ \bibinfo
  {pages} {054610} (\bibinfo {year} {2015})}\BibitemShut {NoStop}%
\bibitem [{\citenamefont {Bulgac}\ \emph {et~al.}(2016)\citenamefont {Bulgac},
  \citenamefont {Magierski}, \citenamefont {Roche},\ and\ \citenamefont
  {Stetcu}}]{Bulgac2016Pu240_fission}%
  \BibitemOpen
  \bibfield  {author} {\bibinfo {author} {\bibfnamefont {A.}~\bibnamefont
  {Bulgac}}, \bibinfo {author} {\bibfnamefont {P.}~\bibnamefont {Magierski}},
  \bibinfo {author} {\bibfnamefont {K.~J.}\ \bibnamefont {Roche}}, \ and\
  \bibinfo {author} {\bibfnamefont {I.}~\bibnamefont {Stetcu}},\ }\href
  {\doibase 10.1103/PhysRevLett.116.122504} {\bibfield  {journal} {\bibinfo
  {journal} {Phys. Rev. Lett.}\ }\textbf {\bibinfo {volume} {116}},\ \bibinfo
  {pages} {122504} (\bibinfo {year} {2016})}\BibitemShut {NoStop}%
\bibitem [{\citenamefont {Tanimura}\ \emph {et~al.}(2017)\citenamefont
  {Tanimura}, \citenamefont {Lacroix},\ and\ \citenamefont
  {Ayik}}]{Tanimura2017fission}%
  \BibitemOpen
  \bibfield  {author} {\bibinfo {author} {\bibfnamefont {Y.}~\bibnamefont
  {Tanimura}}, \bibinfo {author} {\bibfnamefont {D.}~\bibnamefont {Lacroix}}, \
  and\ \bibinfo {author} {\bibfnamefont {S.}~\bibnamefont {Ayik}},\ }\href
  {\doibase 10.1103/PhysRevLett.118.152501} {\bibfield  {journal} {\bibinfo
  {journal} {Phys. Rev. Lett.}\ }\textbf {\bibinfo {volume} {118}},\ \bibinfo
  {pages} {152501} (\bibinfo {year} {2017})}\BibitemShut {NoStop}%
\bibitem [{\citenamefont {Scamps}\ and\ \citenamefont
  {Simenel}(2018)}]{scamps2018impact}%
  \BibitemOpen
  \bibfield  {author} {\bibinfo {author} {\bibfnamefont {G.}~\bibnamefont
  {Scamps}}\ and\ \bibinfo {author} {\bibfnamefont {C.}~\bibnamefont
  {Simenel}},\ }\href {\doibase https://doi.org/10.1038/s41586-018-0780-0}
  {\bibfield  {journal} {\bibinfo  {journal} {Nature}\ }\textbf {\bibinfo
  {volume} {564}},\ \bibinfo {pages} {382} (\bibinfo {year}
  {2018})}\BibitemShut {NoStop}%
\bibitem [{\citenamefont {Guo}\ and\ \citenamefont
  {Nakatsukasa}(2012)}]{Guo2012fusion}%
  \BibitemOpen
  \bibfield  {author} {\bibinfo {author} {\bibfnamefont {L.}~\bibnamefont
  {Guo}}\ and\ \bibinfo {author} {\bibfnamefont {T.}~\bibnamefont
  {Nakatsukasa}},\ }\href {\doibase
  https://doi.org/10.1051/epjconf/20123809003} {\bibfield  {journal} {\bibinfo
  {journal} {EPJ Web Conf.}\ }\textbf {\bibinfo {volume} {38}},\ \bibinfo
  {pages} {09003} (\bibinfo {year} {2012})}\BibitemShut {NoStop}%
\bibitem [{\citenamefont {Umar}\ and\ \citenamefont
  {Oberacker}(2015)}]{Umar2015SHE_TDHF}%
  \BibitemOpen
  \bibfield  {author} {\bibinfo {author} {\bibfnamefont {A.}~\bibnamefont
  {Umar}}\ and\ \bibinfo {author} {\bibfnamefont {V.}~\bibnamefont
  {Oberacker}},\ }\href {\doibase
  https://doi.org/10.1016/j.nuclphysa.2015.02.011} {\bibfield  {journal}
  {\bibinfo  {journal} {Nucl. Phys. A}\ }\textbf {\bibinfo {volume} {944}},\
  \bibinfo {pages} {238} (\bibinfo {year} {2015})}\BibitemShut {NoStop}%
\bibitem [{\citenamefont {Yu}\ and\ \citenamefont {Guo}(2017)}]{Yu2017TDHF3D}%
  \BibitemOpen
  \bibfield  {author} {\bibinfo {author} {\bibfnamefont {C.}~\bibnamefont
  {Yu}}\ and\ \bibinfo {author} {\bibfnamefont {L.}~\bibnamefont {Guo}},\
  }\href {\doibase 10.1007/s11433-017-9063-3} {\bibfield  {journal} {\bibinfo
  {journal} {Sci. China-Phys. Mech. Astron.}\ }\textbf {\bibinfo {volume}
  {60}},\ \bibinfo {pages} {092011} (\bibinfo {year} {2017})}\BibitemShut
  {NoStop}%
\bibitem [{\citenamefont {Guo}\ \emph {et~al.}(2018{\natexlab{a}})\citenamefont
  {Guo}, \citenamefont {Shen}, \citenamefont {Yu},\ and\ \citenamefont
  {Wu}}]{Guo2018TDHF_fusion}%
  \BibitemOpen
  \bibfield  {author} {\bibinfo {author} {\bibfnamefont {L.}~\bibnamefont
  {Guo}}, \bibinfo {author} {\bibfnamefont {C.}~\bibnamefont {Shen}}, \bibinfo
  {author} {\bibfnamefont {C.}~\bibnamefont {Yu}}, \ and\ \bibinfo {author}
  {\bibfnamefont {Z.}~\bibnamefont {Wu}},\ }\href {\doibase
  10.1103/PhysRevC.98.064609} {\bibfield  {journal} {\bibinfo  {journal} {Phys.
  Rev. C}\ }\textbf {\bibinfo {volume} {98}},\ \bibinfo {pages} {064609}
  (\bibinfo {year} {2018}{\natexlab{a}})}\BibitemShut {NoStop}%
\bibitem [{\citenamefont {Guo}\ \emph {et~al.}(2018{\natexlab{b}})\citenamefont
  {Guo}, \citenamefont {Simenel}, \citenamefont {Shi},\ and\ \citenamefont
  {Yu}}]{Guo2018tensor_fusion}%
  \BibitemOpen
  \bibfield  {author} {\bibinfo {author} {\bibfnamefont {L.}~\bibnamefont
  {Guo}}, \bibinfo {author} {\bibfnamefont {C.}~\bibnamefont {Simenel}},
  \bibinfo {author} {\bibfnamefont {L.}~\bibnamefont {Shi}}, \ and\ \bibinfo
  {author} {\bibfnamefont {C.}~\bibnamefont {Yu}},\ }\href {\doibase
  https://doi.org/10.1016/j.physletb.2018.05.066} {\bibfield  {journal}
  {\bibinfo  {journal} {Phys. Lett. B}\ }\textbf {\bibinfo {volume} {782}},\
  \bibinfo {pages} {401} (\bibinfo {year} {2018}{\natexlab{b}})}\BibitemShut
  {NoStop}%
\bibitem [{\citenamefont {Maruhn}\ \emph {et~al.}(2005)\citenamefont {Maruhn},
  \citenamefont {Reinhard}, \citenamefont {Stevenson}, \citenamefont {Stone},\
  and\ \citenamefont {Strayer}}]{maruhn2005TDHF_GDR}%
  \BibitemOpen
  \bibfield  {author} {\bibinfo {author} {\bibfnamefont {J.~A.}\ \bibnamefont
  {Maruhn}}, \bibinfo {author} {\bibfnamefont {P.~G.}\ \bibnamefont
  {Reinhard}}, \bibinfo {author} {\bibfnamefont {P.~D.}\ \bibnamefont
  {Stevenson}}, \bibinfo {author} {\bibfnamefont {J.~R.}\ \bibnamefont
  {Stone}}, \ and\ \bibinfo {author} {\bibfnamefont {M.~R.}\ \bibnamefont
  {Strayer}},\ }\href {\doibase 10.1103/PhysRevC.71.064328} {\bibfield
  {journal} {\bibinfo  {journal} {Phys. Rev. C}\ }\textbf {\bibinfo {volume}
  {71}},\ \bibinfo {pages} {064328} (\bibinfo {year} {2005})}\BibitemShut
  {NoStop}%
\bibitem [{\citenamefont {Reinhard}\ \emph {et~al.}(2007)\citenamefont
  {Reinhard}, \citenamefont {Guo},\ and\ \citenamefont
  {Maruhn}}]{Reinhard2007TDHF_GR}%
  \BibitemOpen
  \bibfield  {author} {\bibinfo {author} {\bibfnamefont {P.~G.}\ \bibnamefont
  {Reinhard}}, \bibinfo {author} {\bibfnamefont {L.}~\bibnamefont {Guo}}, \
  and\ \bibinfo {author} {\bibfnamefont {J.~A.}\ \bibnamefont {Maruhn}},\
  }\href {\doibase 10.1140/epja/i2007-10366-9} {\bibfield  {journal} {\bibinfo
  {journal} {Eur. Phys. J. A}\ }\textbf {\bibinfo {volume} {32}},\ \bibinfo
  {pages} {19} (\bibinfo {year} {2007})}\BibitemShut {NoStop}%
\bibitem [{\citenamefont {Schuetrumpf}\ \emph {et~al.}(2016)\citenamefont
  {Schuetrumpf}, \citenamefont {Nazarewicz},\ and\ \citenamefont
  {Reinhard}}]{Schuetrumpf2016TABC}%
  \BibitemOpen
  \bibfield  {author} {\bibinfo {author} {\bibfnamefont {B.}~\bibnamefont
  {Schuetrumpf}}, \bibinfo {author} {\bibfnamefont {W.}~\bibnamefont
  {Nazarewicz}}, \ and\ \bibinfo {author} {\bibfnamefont {P.-G.}\ \bibnamefont
  {Reinhard}},\ }\href {\doibase 10.1103/PhysRevC.93.054304} {\bibfield
  {journal} {\bibinfo  {journal} {Phys. Rev. C}\ }\textbf {\bibinfo {volume}
  {93}},\ \bibinfo {pages} {054304} (\bibinfo {year} {2016})}\BibitemShut
  {NoStop}%
\bibitem [{\citenamefont {Umar}\ \emph {et~al.}(2010)\citenamefont {Umar},
  \citenamefont {Maruhn}, \citenamefont {Itagaki},\ and\ \citenamefont
  {Oberacker}}]{Umar2010TDHF_C12}%
  \BibitemOpen
  \bibfield  {author} {\bibinfo {author} {\bibfnamefont {A.~S.}\ \bibnamefont
  {Umar}}, \bibinfo {author} {\bibfnamefont {J.~A.}\ \bibnamefont {Maruhn}},
  \bibinfo {author} {\bibfnamefont {N.}~\bibnamefont {Itagaki}}, \ and\
  \bibinfo {author} {\bibfnamefont {V.~E.}\ \bibnamefont {Oberacker}},\ }\href
  {\doibase 10.1103/PhysRevLett.104.212503} {\bibfield  {journal} {\bibinfo
  {journal} {Phys. Rev. Lett.}\ }\textbf {\bibinfo {volume} {104}},\ \bibinfo
  {pages} {212503} (\bibinfo {year} {2010})}\BibitemShut {NoStop}%
\bibitem [{\citenamefont {M\"{u}ller}(1981)}]{MULLER1981TDRMF}%
  \BibitemOpen
  \bibfield  {author} {\bibinfo {author} {\bibfnamefont {K.-H.}\ \bibnamefont
  {M\"{u}ller}},\ }\href {\doibase
  https://doi.org/10.1016/0375-9474(81)90047-6} {\bibfield  {journal} {\bibinfo
   {journal} {Nucl. Phys. A}\ }\textbf {\bibinfo {volume} {372}},\ \bibinfo
  {pages} {459} (\bibinfo {year} {1981})}\BibitemShut {NoStop}%
\bibitem [{\citenamefont {Cusson}\ \emph {et~al.}(1985)\citenamefont {Cusson},
  \citenamefont {Reinhard}, \citenamefont {Molitoris}, \citenamefont
  {St\"ocker}, \citenamefont {Strayer},\ and\ \citenamefont
  {Greiner}}]{Cusson1985TDCDFT}%
  \BibitemOpen
  \bibfield  {author} {\bibinfo {author} {\bibfnamefont {R.~Y.}\ \bibnamefont
  {Cusson}}, \bibinfo {author} {\bibfnamefont {P.~G.}\ \bibnamefont
  {Reinhard}}, \bibinfo {author} {\bibfnamefont {J.~J.}\ \bibnamefont
  {Molitoris}}, \bibinfo {author} {\bibfnamefont {H.}~\bibnamefont
  {St\"ocker}}, \bibinfo {author} {\bibfnamefont {M.~R.}\ \bibnamefont
  {Strayer}}, \ and\ \bibinfo {author} {\bibfnamefont {W.}~\bibnamefont
  {Greiner}},\ }\href {\doibase 10.1103/PhysRevLett.55.2786} {\bibfield
  {journal} {\bibinfo  {journal} {Phys. Rev. Lett.}\ }\textbf {\bibinfo
  {volume} {55}},\ \bibinfo {pages} {2786} (\bibinfo {year}
  {1985})}\BibitemShut {NoStop}%
\bibitem [{\citenamefont {Bai}\ \emph {et~al.}(1987)\citenamefont {Bai},
  \citenamefont {Cusson}, \citenamefont {Wu}, \citenamefont {Reinhard},
  \citenamefont {Stoecker}, \citenamefont {Greiner},\ and\ \citenamefont
  {Strayer}}]{Bai1987TDCDFT}%
  \BibitemOpen
  \bibfield  {author} {\bibinfo {author} {\bibfnamefont {J.~J.}\ \bibnamefont
  {Bai}}, \bibinfo {author} {\bibfnamefont {R.~Y.}\ \bibnamefont {Cusson}},
  \bibinfo {author} {\bibfnamefont {J.}~\bibnamefont {Wu}}, \bibinfo {author}
  {\bibfnamefont {P.~G.}\ \bibnamefont {Reinhard}}, \bibinfo {author}
  {\bibfnamefont {H.}~\bibnamefont {Stoecker}}, \bibinfo {author}
  {\bibfnamefont {W.}~\bibnamefont {Greiner}}, \ and\ \bibinfo {author}
  {\bibfnamefont {M.~R.}\ \bibnamefont {Strayer}},\ }\href {\doibase
  10.1007/BF01297581} {\bibfield  {journal} {\bibinfo  {journal} {Z. Phys. A}\
  }\textbf {\bibinfo {volume} {326}},\ \bibinfo {pages} {269} (\bibinfo {year}
  {1987})}\BibitemShut {NoStop}%
\bibitem [{\citenamefont {Vretenar}\ \emph {et~al.}(1993)\citenamefont
  {Vretenar}, \citenamefont {Berghammer},\ and\ \citenamefont
  {Ring}}]{Vretenar1993TDRMF}%
  \BibitemOpen
  \bibfield  {author} {\bibinfo {author} {\bibfnamefont {D.}~\bibnamefont
  {Vretenar}}, \bibinfo {author} {\bibfnamefont {H.}~\bibnamefont
  {Berghammer}}, \ and\ \bibinfo {author} {\bibfnamefont {P.}~\bibnamefont
  {Ring}},\ }\href {\doibase https://doi.org/10.1016/0370-2693(93)90776-E}
  {\bibfield  {journal} {\bibinfo  {journal} {Phys. Lett. B}\ }\textbf
  {\bibinfo {volume} {319}},\ \bibinfo {pages} {29} (\bibinfo {year}
  {1993})}\BibitemShut {NoStop}%
\bibitem [{\citenamefont {Vretenar}\ \emph {et~al.}(1995)\citenamefont
  {Vretenar}, \citenamefont {Berghammer},\ and\ \citenamefont
  {Ring}}]{VRETENAR1995TDRMF}%
  \BibitemOpen
  \bibfield  {author} {\bibinfo {author} {\bibfnamefont {D.}~\bibnamefont
  {Vretenar}}, \bibinfo {author} {\bibfnamefont {H.}~\bibnamefont
  {Berghammer}}, \ and\ \bibinfo {author} {\bibfnamefont {P.}~\bibnamefont
  {Ring}},\ }\href {\doibase https://doi.org/10.1016/0375-9474(94)00417-L}
  {\bibfield  {journal} {\bibinfo  {journal} {Nucl. Phys. A}\ }\textbf
  {\bibinfo {volume} {581}},\ \bibinfo {pages} {679} (\bibinfo {year}
  {1995})}\BibitemShut {NoStop}%
\bibitem [{\citenamefont {Zhang}\ \emph {et~al.}(2009)\citenamefont {Zhang},
  \citenamefont {Liang},\ and\ \citenamefont {Meng}}]{zhang2009first}%
  \BibitemOpen
  \bibfield  {author} {\bibinfo {author} {\bibfnamefont {Y.}~\bibnamefont
  {Zhang}}, \bibinfo {author} {\bibfnamefont {H.~Z.}\ \bibnamefont {Liang}}, \
  and\ \bibinfo {author} {\bibfnamefont {J.}~\bibnamefont {Meng}},\ }\href
  {http://stacks.iop.org/1674-1137/33/i=S1/a=036} {\bibfield  {journal}
  {\bibinfo  {journal} {Chin. Phys. C}\ }\textbf {\bibinfo {volume} {33}},\
  \bibinfo {pages} {113} (\bibinfo {year} {2009})}\BibitemShut {NoStop}%
\bibitem [{\citenamefont {Zhang}\ \emph {et~al.}(2010)\citenamefont {Zhang},
  \citenamefont {Liang},\ and\ \citenamefont {Meng}}]{ZhangIJMPE2010}%
  \BibitemOpen
  \bibfield  {author} {\bibinfo {author} {\bibfnamefont {Y.}~\bibnamefont
  {Zhang}}, \bibinfo {author} {\bibfnamefont {H.}~\bibnamefont {Liang}}, \ and\
  \bibinfo {author} {\bibfnamefont {J.}~\bibnamefont {Meng}},\ }\href {\doibase
  10.1142/S0218301310014637} {\bibfield  {journal} {\bibinfo  {journal} {Int.
  J. Mod. Phys. E}\ }\textbf {\bibinfo {volume} {19}},\ \bibinfo {pages} {55}
  (\bibinfo {year} {2010})}\BibitemShut {NoStop}%
\bibitem [{\citenamefont {Hagino}\ and\ \citenamefont
  {Tanimura}(2010)}]{hagino2010iterative}%
  \BibitemOpen
  \bibfield  {author} {\bibinfo {author} {\bibfnamefont {K.}~\bibnamefont
  {Hagino}}\ and\ \bibinfo {author} {\bibfnamefont {Y.}~\bibnamefont
  {Tanimura}},\ }\href {\doibase 10.1103/PhysRevC.82.057301} {\bibfield
  {journal} {\bibinfo  {journal} {Phys. Rev. C}\ }\textbf {\bibinfo {volume}
  {82}},\ \bibinfo {pages} {057301} (\bibinfo {year} {2010})}\BibitemShut
  {NoStop}%
\bibitem [{\citenamefont {Tanimura}\ \emph {et~al.}(2015)\citenamefont
  {Tanimura}, \citenamefont {Hagino},\ and\ \citenamefont
  {Liang}}]{tanimura20153d}%
  \BibitemOpen
  \bibfield  {author} {\bibinfo {author} {\bibfnamefont {Y.}~\bibnamefont
  {Tanimura}}, \bibinfo {author} {\bibfnamefont {K.}~\bibnamefont {Hagino}}, \
  and\ \bibinfo {author} {\bibfnamefont {H.~Z.}\ \bibnamefont {Liang}},\ }\href
  {\doibase 10.1093/ptep/ptv083} {\bibfield  {journal} {\bibinfo  {journal}
  {Prog. Theor. Exp. Phys.}\ }\textbf {\bibinfo {volume} {2015}},\ \bibinfo
  {pages} {073D01} (\bibinfo {year} {2015})}\BibitemShut {NoStop}%
\bibitem [{\citenamefont {Ren}\ \emph {et~al.}(2017)\citenamefont {Ren},
  \citenamefont {Zhang},\ and\ \citenamefont {Meng}}]{REN2017Dirac3D}%
  \BibitemOpen
  \bibfield  {author} {\bibinfo {author} {\bibfnamefont {Z.~X.}\ \bibnamefont
  {Ren}}, \bibinfo {author} {\bibfnamefont {S.~Q.}\ \bibnamefont {Zhang}}, \
  and\ \bibinfo {author} {\bibfnamefont {J.}~\bibnamefont {Meng}},\ }\href
  {\doibase 10.1103/PhysRevC.95.024313} {\bibfield  {journal} {\bibinfo
  {journal} {Phys. Rev. C}\ }\textbf {\bibinfo {volume} {95}},\ \bibinfo
  {pages} {024313} (\bibinfo {year} {2017})}\BibitemShut {NoStop}%
\bibitem [{\citenamefont {Ren}\ \emph {et~al.}(2019)\citenamefont {Ren},
  \citenamefont {Zhang}, \citenamefont {Zhao}, \citenamefont {Itagaki},
  \citenamefont {Maruhn},\ and\ \citenamefont {Meng}}]{Ren2019C12LCS}%
  \BibitemOpen
  \bibfield  {author} {\bibinfo {author} {\bibfnamefont {Z.~X.}\ \bibnamefont
  {Ren}}, \bibinfo {author} {\bibfnamefont {S.~Q.}\ \bibnamefont {Zhang}},
  \bibinfo {author} {\bibfnamefont {P.~W.}\ \bibnamefont {Zhao}}, \bibinfo
  {author} {\bibfnamefont {N.}~\bibnamefont {Itagaki}}, \bibinfo {author}
  {\bibfnamefont {J.~A.}\ \bibnamefont {Maruhn}}, \ and\ \bibinfo {author}
  {\bibfnamefont {J.}~\bibnamefont {Meng}},\ }\href {\doibase
  https://doi.org/10.1007/s11433-019-9412-3} {\bibfield  {journal} {\bibinfo
  {journal} {Sci. China-Phys. Mech. Astron.}\ }\textbf {\bibinfo {volume}
  {62}},\ \bibinfo {pages} {112062} (\bibinfo {year} {2019})}\BibitemShut
  {NoStop}%
\bibitem [{\citenamefont {Ren}\ \emph {et~al.}(2020{\natexlab{a}})\citenamefont
  {Ren}, \citenamefont {Zhao}, \citenamefont {Zhang},\ and\ \citenamefont
  {Meng}}]{REN2020Toroidal}%
  \BibitemOpen
  \bibfield  {author} {\bibinfo {author} {\bibfnamefont {Z.~X.}\ \bibnamefont
  {Ren}}, \bibinfo {author} {\bibfnamefont {P.~W.}\ \bibnamefont {Zhao}},
  \bibinfo {author} {\bibfnamefont {S.~Q.}\ \bibnamefont {Zhang}}, \ and\
  \bibinfo {author} {\bibfnamefont {J.}~\bibnamefont {Meng}},\ }\href {\doibase
  https://doi.org/10.1016/j.nuclphysa.2020.121696} {\bibfield  {journal}
  {\bibinfo  {journal} {Nucl. Phys. A}\ }\textbf {\bibinfo {volume} {996}},\
  \bibinfo {pages} {121696} (\bibinfo {year} {2020}{\natexlab{a}})}\BibitemShut
  {NoStop}%
\bibitem [{\citenamefont {Ren}\ \emph {et~al.}(2020{\natexlab{b}})\citenamefont
  {Ren}, \citenamefont {Zhao},\ and\ \citenamefont {Meng}}]{Ren2020HeBeTDCDFT}%
  \BibitemOpen
  \bibfield  {author} {\bibinfo {author} {\bibfnamefont {Z.~X.}\ \bibnamefont
  {Ren}}, \bibinfo {author} {\bibfnamefont {P.~W.}\ \bibnamefont {Zhao}}, \
  and\ \bibinfo {author} {\bibfnamefont {J.}~\bibnamefont {Meng}},\ }\href
  {\doibase https://doi.org/10.1016/j.physletb.2019.135194} {\bibfield
  {journal} {\bibinfo  {journal} {Phys. Lett. B}\ }\textbf {\bibinfo {volume}
  {801}},\ \bibinfo {pages} {135194} (\bibinfo {year}
  {2020}{\natexlab{b}})}\BibitemShut {NoStop}%
\bibitem [{\citenamefont {Zhao}\ \emph {et~al.}(2010)\citenamefont {Zhao},
  \citenamefont {Li}, \citenamefont {Yao},\ and\ \citenamefont
  {Meng}}]{ZhaoPC-PK1}%
  \BibitemOpen
  \bibfield  {author} {\bibinfo {author} {\bibfnamefont {P.~W.}\ \bibnamefont
  {Zhao}}, \bibinfo {author} {\bibfnamefont {Z.~P.}\ \bibnamefont {Li}},
  \bibinfo {author} {\bibfnamefont {J.~M.}\ \bibnamefont {Yao}}, \ and\
  \bibinfo {author} {\bibfnamefont {J.}~\bibnamefont {Meng}},\ }\href {\doibase
  10.1103/PhysRevC.82.054319} {\bibfield  {journal} {\bibinfo  {journal} {Phys.
  Rev. C}\ }\textbf {\bibinfo {volume} {82}},\ \bibinfo {pages} {054319}
  (\bibinfo {year} {2010})}\BibitemShut {NoStop}%
\bibitem [{\citenamefont {van Leeuwen}(1999)}]{Leeuwen1999TDDFT}%
  \BibitemOpen
  \bibfield  {author} {\bibinfo {author} {\bibfnamefont {R.}~\bibnamefont {van
  Leeuwen}},\ }\href {\doibase 10.1103/PhysRevLett.82.3863} {\bibfield
  {journal} {\bibinfo  {journal} {Phys. Rev. Lett.}\ }\textbf {\bibinfo
  {volume} {82}},\ \bibinfo {pages} {3863} (\bibinfo {year}
  {1999})}\BibitemShut {NoStop}%
\bibitem [{\citenamefont {Greiner}(2013)}]{greiner2013relativistic}%
  \BibitemOpen
  \bibfield  {author} {\bibinfo {author} {\bibfnamefont {W.}~\bibnamefont
  {Greiner}},\ }\href@noop {} {\emph {\bibinfo {title} {Relativistic Quantum
  Mechanics: Wave Equations}}}\ (\bibinfo  {publisher} {Springer Science \&
  Business Media},\ \bibinfo {year} {2013})\BibitemShut {NoStop}%
\bibitem [{\citenamefont {Maruhn}\ \emph {et~al.}(2014)\citenamefont {Maruhn},
  \citenamefont {Reinhard}, \citenamefont {Stevenson},\ and\ \citenamefont
  {Umar}}]{Maruhn2014CPC}%
  \BibitemOpen
  \bibfield  {author} {\bibinfo {author} {\bibfnamefont {J.~A.}\ \bibnamefont
  {Maruhn}}, \bibinfo {author} {\bibfnamefont {P.~G.}\ \bibnamefont
  {Reinhard}}, \bibinfo {author} {\bibfnamefont {P.~D.}\ \bibnamefont
  {Stevenson}}, \ and\ \bibinfo {author} {\bibfnamefont {A.~S.}\ \bibnamefont
  {Umar}},\ }\href {\doibase http://dx.doi.org/10.1016/j.cpc.2014.04.008}
  {\bibfield  {journal} {\bibinfo  {journal} {Comput. Phys. Commun.}\ }\textbf
  {\bibinfo {volume} {185}},\ \bibinfo {pages} {2195} (\bibinfo {year}
  {2014})}\BibitemShut {NoStop}%
\bibitem [{\citenamefont {Eastwood}\ and\ \citenamefont
  {Brownrigg}(1979)}]{eastwood1979remarks}%
  \BibitemOpen
  \bibfield  {author} {\bibinfo {author} {\bibfnamefont {J.}~\bibnamefont
  {Eastwood}}\ and\ \bibinfo {author} {\bibfnamefont {D.}~\bibnamefont
  {Brownrigg}},\ }\href {\doibase https://doi.org/10.1016/0021-9991(79)90139-6}
  {\bibfield  {journal} {\bibinfo  {journal} {J. Comput. Phys.}\ }\textbf
  {\bibinfo {volume} {32}},\ \bibinfo {pages} {24} (\bibinfo {year}
  {1979})}\BibitemShut {NoStop}%
\bibitem [{\citenamefont {Ring}\ and\ \citenamefont
  {Schuck}(2004)}]{ring2004nuclear}%
  \BibitemOpen
  \bibfield  {author} {\bibinfo {author} {\bibfnamefont {P.}~\bibnamefont
  {Ring}}\ and\ \bibinfo {author} {\bibfnamefont {P.}~\bibnamefont {Schuck}},\
  }\href@noop {} {\emph {\bibinfo {title} {The nuclear many-body problem}}}\
  (\bibinfo  {publisher} {Springer Science \& Business Media},\ \bibinfo {year}
  {2004})\BibitemShut {NoStop}%
\bibitem [{\citenamefont {Dai}\ \emph {et~al.}(2014)\citenamefont {Dai},
  \citenamefont {Guo}, \citenamefont {Zhao},\ and\ \citenamefont
  {Zhou}}]{Dai2014Dissipation}%
  \BibitemOpen
  \bibfield  {author} {\bibinfo {author} {\bibfnamefont {G.-F.}\ \bibnamefont
  {Dai}}, \bibinfo {author} {\bibfnamefont {L.}~\bibnamefont {Guo}}, \bibinfo
  {author} {\bibfnamefont {E.-G.}\ \bibnamefont {Zhao}}, \ and\ \bibinfo
  {author} {\bibfnamefont {S.-G.}\ \bibnamefont {Zhou}},\ }\href {\doibase
  10.1103/PhysRevC.90.044609} {\bibfield  {journal} {\bibinfo  {journal} {Phys.
  Rev. C}\ }\textbf {\bibinfo {volume} {90}},\ \bibinfo {pages} {044609}
  (\bibinfo {year} {2014})}\BibitemShut {NoStop}%
\bibitem [{\citenamefont {Fernandez}\ \emph {et~al.}(1978)\citenamefont
  {Fernandez}, \citenamefont {Gaarde}, \citenamefont {Larsen}, \citenamefont
  {Pontoppidan},\ and\ \citenamefont {Videbaek}}]{Fernandez1978O16fusion}%
  \BibitemOpen
  \bibfield  {author} {\bibinfo {author} {\bibfnamefont {B.}~\bibnamefont
  {Fernandez}}, \bibinfo {author} {\bibfnamefont {C.}~\bibnamefont {Gaarde}},
  \bibinfo {author} {\bibfnamefont {J.}~\bibnamefont {Larsen}}, \bibinfo
  {author} {\bibfnamefont {S.}~\bibnamefont {Pontoppidan}}, \ and\ \bibinfo
  {author} {\bibfnamefont {F.}~\bibnamefont {Videbaek}},\ }\href {\doibase
  https://doi.org/10.1016/0375-9474(78)90327-5} {\bibfield  {journal} {\bibinfo
   {journal} {Nucl. Phys. A}\ }\textbf {\bibinfo {volume} {306}},\ \bibinfo
  {pages} {259} (\bibinfo {year} {1978})}\BibitemShut {NoStop}%
\bibitem [{\citenamefont {Tserruya}\ \emph {et~al.}(1978)\citenamefont
  {Tserruya}, \citenamefont {Eisen}, \citenamefont {Pelte}, \citenamefont
  {Gavron}, \citenamefont {Oeschler}, \citenamefont {Berndt},\ and\
  \citenamefont {Harney}}]{Tserruya1978O16fusion}%
  \BibitemOpen
  \bibfield  {author} {\bibinfo {author} {\bibfnamefont {I.}~\bibnamefont
  {Tserruya}}, \bibinfo {author} {\bibfnamefont {Y.}~\bibnamefont {Eisen}},
  \bibinfo {author} {\bibfnamefont {D.}~\bibnamefont {Pelte}}, \bibinfo
  {author} {\bibfnamefont {A.}~\bibnamefont {Gavron}}, \bibinfo {author}
  {\bibfnamefont {H.}~\bibnamefont {Oeschler}}, \bibinfo {author}
  {\bibfnamefont {D.}~\bibnamefont {Berndt}}, \ and\ \bibinfo {author}
  {\bibfnamefont {H.~L.}\ \bibnamefont {Harney}},\ }\href {\doibase
  10.1103/PhysRevC.18.1688} {\bibfield  {journal} {\bibinfo  {journal} {Phys.
  Rev. C}\ }\textbf {\bibinfo {volume} {18}},\ \bibinfo {pages} {1688}
  (\bibinfo {year} {1978})}\BibitemShut {NoStop}%
\bibitem [{\citenamefont {Kolata}\ \emph {et~al.}(1979)\citenamefont {Kolata},
  \citenamefont {Freeman}, \citenamefont {Haas}, \citenamefont {Heusch},\ and\
  \citenamefont {Gallmann}}]{Kolata1979O16fusion}%
  \BibitemOpen
  \bibfield  {author} {\bibinfo {author} {\bibfnamefont {J.~J.}\ \bibnamefont
  {Kolata}}, \bibinfo {author} {\bibfnamefont {R.~M.}\ \bibnamefont {Freeman}},
  \bibinfo {author} {\bibfnamefont {F.}~\bibnamefont {Haas}}, \bibinfo {author}
  {\bibfnamefont {B.}~\bibnamefont {Heusch}}, \ and\ \bibinfo {author}
  {\bibfnamefont {A.}~\bibnamefont {Gallmann}},\ }\href {\doibase
  10.1103/PhysRevC.19.2237} {\bibfield  {journal} {\bibinfo  {journal} {Phys.
  Rev. C}\ }\textbf {\bibinfo {volume} {19}},\ \bibinfo {pages} {2237}
  (\bibinfo {year} {1979})}\BibitemShut {NoStop}%
\bibitem [{\citenamefont {Wu}\ and\ \citenamefont
  {Barnes}(1984)}]{Wu1984O16fusion}%
  \BibitemOpen
  \bibfield  {author} {\bibinfo {author} {\bibfnamefont {S.-C.}\ \bibnamefont
  {Wu}}\ and\ \bibinfo {author} {\bibfnamefont {C.}~\bibnamefont {Barnes}},\
  }\href {\doibase https://doi.org/10.1016/0375-9474(84)90523-2} {\bibfield
  {journal} {\bibinfo  {journal} {Nucl. Phys. A}\ }\textbf {\bibinfo {volume}
  {422}},\ \bibinfo {pages} {373} (\bibinfo {year} {1984})}\BibitemShut
  {NoStop}%
\bibitem [{\citenamefont {Thomas}\ \emph {et~al.}(1986)\citenamefont {Thomas},
  \citenamefont {Chen}, \citenamefont {Hinds}, \citenamefont {Meredith},\ and\
  \citenamefont {Olson}}]{Thomas1986O16O16fusion}%
  \BibitemOpen
  \bibfield  {author} {\bibinfo {author} {\bibfnamefont {J.}~\bibnamefont
  {Thomas}}, \bibinfo {author} {\bibfnamefont {Y.~T.}\ \bibnamefont {Chen}},
  \bibinfo {author} {\bibfnamefont {S.}~\bibnamefont {Hinds}}, \bibinfo
  {author} {\bibfnamefont {D.}~\bibnamefont {Meredith}}, \ and\ \bibinfo
  {author} {\bibfnamefont {M.}~\bibnamefont {Olson}},\ }\href {\doibase
  10.1103/PhysRevC.33.1679} {\bibfield  {journal} {\bibinfo  {journal} {Phys.
  Rev. C}\ }\textbf {\bibinfo {volume} {33}},\ \bibinfo {pages} {1679}
  (\bibinfo {year} {1986})}\BibitemShut {NoStop}%
\bibitem [{\citenamefont {Umar}\ \emph {et~al.}(1986)\citenamefont {Umar},
  \citenamefont {Strayer},\ and\ \citenamefont {Reinhard}}]{Umar1986TDHFLS}%
  \BibitemOpen
  \bibfield  {author} {\bibinfo {author} {\bibfnamefont {A.~S.}\ \bibnamefont
  {Umar}}, \bibinfo {author} {\bibfnamefont {M.~R.}\ \bibnamefont {Strayer}}, \
  and\ \bibinfo {author} {\bibfnamefont {P.~G.}\ \bibnamefont {Reinhard}},\
  }\href {\doibase 10.1103/PhysRevLett.56.2793} {\bibfield  {journal} {\bibinfo
   {journal} {Phys. Rev. Lett.}\ }\textbf {\bibinfo {volume} {56}},\ \bibinfo
  {pages} {2793} (\bibinfo {year} {1986})}\BibitemShut {NoStop}%
\bibitem [{\citenamefont {Reinhard}\ \emph {et~al.}(1988)\citenamefont
  {Reinhard}, \citenamefont {Umar}, \citenamefont {Davies}, \citenamefont
  {Strayer},\ and\ \citenamefont {Lee}}]{Reinhard1988TDHFLS}%
  \BibitemOpen
  \bibfield  {author} {\bibinfo {author} {\bibfnamefont {P.-G.}\ \bibnamefont
  {Reinhard}}, \bibinfo {author} {\bibfnamefont {A.~S.}\ \bibnamefont {Umar}},
  \bibinfo {author} {\bibfnamefont {K.~T.~R.}\ \bibnamefont {Davies}}, \bibinfo
  {author} {\bibfnamefont {M.~R.}\ \bibnamefont {Strayer}}, \ and\ \bibinfo
  {author} {\bibfnamefont {S.-J.}\ \bibnamefont {Lee}},\ }\href {\doibase
  10.1103/PhysRevC.37.1026} {\bibfield  {journal} {\bibinfo  {journal} {Phys.
  Rev. C}\ }\textbf {\bibinfo {volume} {37}},\ \bibinfo {pages} {1026}
  (\bibinfo {year} {1988})}\BibitemShut {NoStop}%
\bibitem [{\citenamefont {Hill}\ and\ \citenamefont
  {Wheeler}(1953)}]{Hill1953PhysicalReview}%
  \BibitemOpen
  \bibfield  {author} {\bibinfo {author} {\bibfnamefont {D.~L.}\ \bibnamefont
  {Hill}}\ and\ \bibinfo {author} {\bibfnamefont {J.~A.}\ \bibnamefont
  {Wheeler}},\ }\href {\doibase 10.1103/PhysRev.89.1102} {\bibfield  {journal}
  {\bibinfo  {journal} {Phys. Rev.}\ }\textbf {\bibinfo {volume} {89}},\
  \bibinfo {pages} {1102} (\bibinfo {year} {1953})}\BibitemShut {NoStop}%
\bibitem [{\citenamefont {Esbensen}(2012)}]{Esbensen2012HighEdata}%
  \BibitemOpen
  \bibfield  {author} {\bibinfo {author} {\bibfnamefont {H.}~\bibnamefont
  {Esbensen}},\ }\href {\doibase 10.1103/PhysRevC.85.064611} {\bibfield
  {journal} {\bibinfo  {journal} {Phys. Rev. C}\ }\textbf {\bibinfo {volume}
  {85}},\ \bibinfo {pages} {064611} (\bibinfo {year} {2012})}\BibitemShut
  {NoStop}%
\bibitem [{\citenamefont {Simenel}\ \emph {et~al.}(2013)\citenamefont
  {Simenel}, \citenamefont {Keser}, \citenamefont {Umar},\ and\ \citenamefont
  {Oberacker}}]{Simenel2013O16O16}%
  \BibitemOpen
  \bibfield  {author} {\bibinfo {author} {\bibfnamefont {C.}~\bibnamefont
  {Simenel}}, \bibinfo {author} {\bibfnamefont {R.}~\bibnamefont {Keser}},
  \bibinfo {author} {\bibfnamefont {A.~S.}\ \bibnamefont {Umar}}, \ and\
  \bibinfo {author} {\bibfnamefont {V.~E.}\ \bibnamefont {Oberacker}},\ }\href
  {\doibase 10.1103/PhysRevC.88.024617} {\bibfield  {journal} {\bibinfo
  {journal} {Phys. Rev. C}\ }\textbf {\bibinfo {volume} {88}},\ \bibinfo
  {pages} {024617} (\bibinfo {year} {2013})}\BibitemShut {NoStop}%
\end{thebibliography}

%

\end{document}